\documentclass{aa}  

\usepackage{graphicx}
\usepackage{txfonts}
\usepackage{hyperref}
\hypersetup{colorlinks=true,linkcolor=blue,citecolor=blue,filecolor=blue,urlcolor=blue}

\usepackage{natbib} 
\bibpunct{(}{)}{;}{a}{}{,} 

\begin{document}

   \title{Uncertainties in asteroseismic grid-based estimates of the ages of halo stars}

   \subtitle{}

   \author{S. Moser\inst{1}, G. Valle \inst{1}, M. Dell'Omodarme \inst{1}, S. Degl'Innocenti
     \inst{1, 2}, P.G. Prada Moroni \inst{1,2} 
          }

   \authorrunning{Moser, S. et al.}

   \institute{
Dipartimento di Fisica ``Enrico Fermi'',
Universit\`a di Pisa, Largo Pontecorvo 3, I-56127, Pisa, Italy
\and
 INFN,
 Sezione di Pisa, Largo Pontecorvo 3, I-56127, Pisa, Italy
 }

   \offprints{G. Valle, valle@df.unipi.it}

   \date{Received 29/04/2022; accepted 18/01/2023}

  \abstract
  {
Stellar age determinations for field stars are crucial for studying the evolutionary history of the  Galaxy. The vast majority of the research in this area has so far been focused on stars with typical disk characteristics.
}
   {  
Nowadays, the availability of high-quality asteroseismic data for stars with typical halo characteristics makes  it possible to extend such investigations.
The aim of this paper is to study the precision and theoretical biases in the age determinations of halo stars adopting both asteroseismic and classic observational constraints. 
}
{  
We adopt the well-tested SCEPtER pipeline, covering evolutionary phases up to the red giant branch (RGB). The fitting grids contain stars with mass in the range of 
[0.7; 1.0] $M_{\sun}$  and metallicity [Fe/H] from -2.5 to -0.5, which are typical ranges seen  in the halo population. We investigate several scenarios characterised by different adopted observational uncertainties. We also assess the impact of systematic discrepancies between the recovery grid models and target stars by computing several synthetic grids of stellar models with perturbed input physics.
 }
  {
We achieve more precise asteroseismic age estimates  for old metal-poor stars than for more metallic
stars. In our reference scenario (errors in $\Delta \nu$ and $\nu_{\rm max}$ of 2.5\% and 5\% respectively), we recover ages for stars in the main sequence (MS) or subgiant branch (SGB) with a typical 10\%--20\% precision, while we recover those of RGB stars with a precision of about 60\%. However, recent observations allow  tighter constraints on asteroseismic parameters by about a factor of 3. With this assumption, the age precision in RGB improved to 20\%, while few modifications occur in the other analysed evolutionary phases.
Our investigation of the relevance of systematic discrepancies between grid models and target stars shows that a mismatch in the mixing-length parameter value between grids and targets (from 1.9 to 1.74)
leads to significant bias in the age estimations for MS stars (about 10\%), but this bias is smaller for  SGB and RGB stars. Neglecting the microscopic diffusion effect in the recovery grid leads to a typical 40\% bias in age estimates for stars on the MS.

Finally, we applied the age estimation technique to stars in globular clusters, adopting typical observational uncertainties from the literature. We find a precision in age estimates of around 20\% for MS stars and up to 40\% for RGB stars. These uncertainties are greater than those obtained with classical methods, which are therefore still to be preferred. We also applied the SCEPtER pipeline to the age determination of the stars of the cluster M4, relying on asteroseismic data for seven RGB stars from the literature. 
We obtain a cluster age of $11.9 \pm 1.5$ Gyr and  a mass at the turn-of off $0.86 \pm 0.04$ $M_{\sun}$, which are in good agreement with literature results.
}
{}


   \maketitle

\section{Introduction}\label{sec:intro}

Accurate stellar age estimations are fundamental for recovery of the evolutionary history of the Galaxy.  
However, estimation of  ages of field stars is notoriously difficult because of several uncertainties both in the stellar models used to date stars and in observations \citep[see][for a review]{Soderblom2010}. 
As a consequence,  the precision  of age determinations obtained from the comparison between theory and observations for classical observables is generally poor, with errors of greater than 40\% in several cases  \citep[e.g.][]{jorgensen2005, Takeda07, Soderblom2010, Sanders2018}.
The problem  is even more severe for old, distant stars in the Galactic halo, making  the exploration of the early Galactic history problematic \citep[see e.g.][]{Jofre2011, Guo2016, Das2020, Matsuno2021}.

The recent development of precision asteroseismology thanks to
satellite missions such as CoRoT \citep{Appourchaux2008, Michel2008, Baglin2009},  {\it Kepler}
\citep{Borucki2010, Gilliland2010} and TESS \citep{Ricker2015} has led to noticeable
improvement in stellar age estimations.
Several analyses in the literature \citep[e.g.][]{Gai2011, Chaplin2014, Casagrande2014, eta} have shown that it is possible to achieve age estimations with an average precision of  10\%--20\%, depending on the  evolutionary phase of the target star.

The vast majority of research has been focused on stars near the Sun with typical disk characteristics, simply because most of the available asteroseismic data  covers such targets.
It is only recently that asteroseismology has been used to investigate  the  ages of halo stars  \citep[e.g.][]{Montalban2021, Matsuno2021, Grunblatt2021}.

The aim of the present paper is to evaluate the typical uncertainties and biases in age determinations for field stars with characteristics typical of the halo of the Milky Way. The  analysis presented here was  conducted using targets for which both global asteroseismic quantities (i.e. the average large frequency spacing and the frequency of maximum oscillation power) and classical observables (effective temperature and [Fe/H]) were available. The work addresses both the theoretical assessment of expected errors and the exploration of possible sources of bias---- due to uncertainty in the  input physics of stellar models--- and includes applications to real data.
We closely follow the methodology presented in \citet{eta}, where a similar analysis was performed for stars with higher metallicities and covering a wider age range. This  allows a direct comparison of the results, thus highlighting the differences in age estimation caused by the  difference in the explored metallicity range.
Finally, we compare the precision of our asteroseismic age determinations against those from the isochrone fitting method for the nearby globular cluster (GC) M4, which is the only GC for which asteroseismic data are available.

The paper is organised as follows. In Sect.~\ref{sec:method}, we discuss the method and the grids adopted in our estimation process. Sections~\ref{sec:results} and \ref{sec:errorprop} contain the results and an  investigation of the possible  sources of bias. Section~\ref{sec:M4} addresses the M4 fit. Some concluding remarks are provided in Sect.~\ref{sec:conclusions}.

\section{Grid-based recovery technique}\label{sec:method}

\subsection{Estimation method}\label{Scepter_explain}

We adopted the SCEPtER scheme\footnote{An R library providing the estimation code and grid is available at CRAN: \url{http://CRAN.R-project.org/package=SCEPtER}.} , which is extensively 
 described in \citet{scepter1}.  
For convenience, we summarise the general aspects of the procedure.
Taking $\cal
 S$ to be a star for which the following vector of observed quantities
is available: $q^{\cal S} \equiv \{T_{\rm eff, \cal S}, {\rm [Fe/H]}_{\cal S},
\Delta \nu_{\cal S}, \nu_{\rm max, \cal S}\}$, we let $\sigma = \{\sigma(T_{\rm
  eff, \cal S}), \sigma({\rm [Fe/H]}_{\cal S}), \sigma(\Delta \nu_{\cal S}),
\sigma(\nu_{\rm max, \cal S})\}$ be the nominal uncertainty in the observed
quantities. For each point $j$ on the estimation grid of stellar models, 
we define $q^{j} \equiv \{T_{{\rm eff}, j}, {\rm [Fe/H]}_{j}, \Delta \nu_{j},
\nu_{{\rm max}, j}\}$. 
We set $ {\cal L}_j $ as the likelihood function, defined as
\begin{equation}
{\cal L}_j = \left( \prod_{i=1}^4 \frac{1}{\sqrt{2 \pi} \sigma_i} \right)
\times \exp \left( -\frac{d_j^2}{2} \right),
\label{eq:lik}
\end{equation}
where
\begin{equation}
d_j =  \left\lVert \frac{q^{\cal S} - \tilde q^j}{\sigma} \right\rVert
.\end{equation}

The likelihood function is evaluated for each grid point within $3 \sigma$ of
all the variables from $\cal S$. We let ${\cal L}_{\rm max}$ be the maximum value
obtained in this step. The estimated stellar mass, radius,
and age are obtained
by averaging the corresponding quantity of all the models with a likelihood
of greater than $0.95 \times {\cal L}_{\rm max}$.

The technique can also be employed to construct a Monte Carlo confidence
interval for stellar parameters. 
 To this purpose, a synthetic sample of $n = 10\,000$ stars is
generated, following a multivariate normal distribution with a vector of the mean
$q^{\cal S}$ and a covariance matrix $\Sigma = {\rm diag}(\sigma)$. The medians of the stellar parameters of the $n$
objects  are taken as the best estimates of the true values, and
the 16th and 84th quantiles are adopted as a $1 \sigma$
confidence interval.

\subsection{Standard  grid of stellar models}\label{standard_grid}

The standard estimation grid of stellar models was computed using the FRANEC
stellar evolution code \citep{scilla2008} in the same
configuration that was adopted to compute the Pisa Stellar
Evolution Data Base\footnote{\url{http://astro.df.unipi.it/stellar-models/}} 
for low-mass stars \citep{database2012, stellar}.
The resulting grid consists of 572\,880 points (880 points for 651 evolutionary tracks),
corresponding to evolutionary stages from the ZAMS to the helium flash. Models were computed for masses in the range [0.70; 1.00] $M_{\sun}$ with a step of 0.01 $M_{\sun}$.
This mass range was chosen to include the halo population of the Milky Way, which shows ages generally older than 10 Gyr; it is therefore expected that stars above $\sim 1$ $M_{\sun}$ have already gone past the red giant branch (RGB) and are therefore no longer within the evolutionary phases of interest for this work.
The upper limit of 1.0 $M_{\sun}$ was imposed  to mitigate the impact of the edge effects, which are discussed below. Stars with $M < 0.7$ $M_{\sun}$ and therefore likely still in the main sequence (MS) were not included.  While there are objects in this mass range for which asteroseismic observations are available, their study is beyond the scope of the present paper.
The initial metallicity [Fe/H] was assumed to be in the range [$-2.5$; $-0.5$], with a step of 0.1 dex in order to include the majority of the halo stars.
We did not theoretically investigate the effects of uncertainties in $\alpha$ enhancement, and we assumed the solar-scaled heavy-element mixture presented by \citet{AGSS09} for both grid models and synthetic observations.
This assumption was dropped when estimating the age of the GC M4 in Sect.~\ref{sec:M4}, where we assumed an $\alpha$ enhancement of 0.4 \citep{Marino2008}.
The initial helium abundance was obtained using the generally adopted linear relation for the helium to metal enrichment ratio $Y = Y_p+\frac{\Delta Y}{\Delta Z} Z$ with a primordial $^4$He abundance value of $Y_p = 0.2471$ from \citet{Planck2020}, and  $\Delta
Y/\Delta Z = 2.03$ \citep{Tognelli2021}. The models were computed following the mixing length formalism for convective envelopes assuming a mixing-length parameter $\alpha_{\rm ml} = 1.90$, as obtained from a calibration on the M4 photometric data, in particular in the RGB region (see Fig.~\ref{fig:Iso12}).
Atomic diffusion was included adopting the coefficients given by
\citet{thoul94} for gravitational settling and thermal diffusion. 
To prevent the surface helium and metal depletion for stars
without a convective envelope, a diffusion inhibition mechanism similar to the one discussed in \citet{Chaboyer2001} is adopted.
For the outermost 1\% of the mass of the star, the diffusion velocities were
multiplied by a suppression parabolic factor that takes a value of 1 for 99\% of the mass of the structure and 0 at the base of the atmosphere.
Further details about the input physics adopted in the computations are available in \citet{scepter1, cefeidi}.  

The average large frequency spacing $\Delta \nu$ and
the frequency of maximum  oscillation power $\nu_{\rm max}$ were obtained using a simple scaling from the solar values \citep{Ulrich1986, Brown1991, Kjeldsen1995}: 
\begin{eqnarray}\label{eq:dni}
\frac{\Delta \nu}{\Delta \nu_{\sun}} & = &
\sqrt{\frac{M/M_{\sun}}{(R/R_{\sun})^3}} \quad ,\\  \frac{\nu_{\rm
    max}}{\nu_{\rm max, \sun}} & = & \frac{{M/M_{\sun}}}{ (R/R_{\sun})^2
  \sqrt{ T_{\rm eff}/T_{\rm eff, \sun}} }. \label{eq:nimax}
\end{eqnarray}

The validity of these scaling relations in the RGB phase has been questioned in recent years \citep[e.g.][]{Epstein2014b, Gaulme2016, Viani2017} and using them to fit real observational RGB stars can lead to systematic biases. Corrections of these scaling relations have been investigated by many authors in different ranges of mass, metallicity, and evolutionary phase \citep[e.g.][]{White2011, Sharma2016}.
 \citet{Rodrigued2017} used theoretical models to determine corrections of the reference values $\Delta\nu_{\rm ref}$ to be used instead of the solar large-frequency separation, $\Delta\nu_{\sun}$, as a function of mass, metallicity, and $\nu_{\rm max}$.
These latter authors showed that for RGB stars, with metallicity inside the range $-0.75 \leq {\rm [Fe/H]} \leq -1.00$, mass in the range of $0.8 \leq M_{\sun} \leq 1.0,$ and $\nu_{\rm max}$ within $10 \leq \nu_{\rm max} \leq 40$ $\mu$Hz, the correction at the $\Delta\nu_{\rm ref}$ value to be applied is about -5\%. This result was confirmed by \citet{Tail02022}   who performed an analysis of RGB stars in M4, obtaining a correction from  -3\% in the initial part of the RGB to -5\% for stars near the RGB bump.
We adopted this correction when estimating stellar parameters for stars in M4\footnote{The correction was not applied to the grid, but to the observables to make the procedure faster and easier.} (Sect.~\ref{sec:M4}).
The reliability of the scaling relation is instead of minor relevance when testing the  accuracy of the internal grid and possible sources of bias, because we use exactly the same scaling relations to compute $\Delta \nu$ and $\nu_{\rm max}$ in both the artificial stars and the models. Therefore, no correction was applied in these cases.

\section{Age estimates}\label{sec:results}

The age-recovery procedure was first tested on a synthetic dataset obtained
by sampling $N = 30\,000$ artificial stars from the same standard estimation
grid of stellar models used in the recovery procedure itself.
For each synthetic object, a Gaussian noise was added to the observable quantities to simulate the effects of typical uncertainties on the observations.
In this work, we were interested in stars with characteristics typical of the Milky Way halo, and so we only sampled stars between 9.5 Gyr and 13.8 Gyr.
The lower limit was established from the belief that the vast majority of halo stars are older than 10 Gyr \citep{Jofre2011}, whereas the upper limit comes directly from recent estimations of the  age of the Universe \citep{Planck2020}.

We separately analysed stars in different evolutionary phases: MS, subgiant branch (SGB), and RGB.
Stars evolved past the RGB were not considered because of the necessity to include the uncertainty of the mass-loss efficiency  in calculations; this would require a much extended model grid, significantly increasing the computational times.
The end of the MS was identified as the point at which the central hydrogen abundance drops below $10^{-15}$. The start of the RGB was chosen according to geometrical considerations, finding the point at the SGB end where the tangent to the track in the $\log L - \log T_{\rm eff}$ plane is parallel to the line passing through the RGB bump and the turn-off (TO; identified as the point with maximum effective temperature). 
These choices allow a neat synchronisation of the  evolutionary stages of the tracks.

The synthetic sample contains an equal number of models from each analysed evolutionary phase.
To obtain a more realistic population, stars were sampled adopting a typical initial mass function distribution filter \citep[e.g.][]{Salpeter1955, Kroupa2002}.
Due to the selection restrictions, the synthetic sample does not cover the whole mass range of the estimation grid, because stars older than 10 Gyr and more massive than $\sim 1.0$ $M_{\sun}$ have already evolved past the RGB phase. For the same reason, the mass range of interest differs for each investigated evolutionary phase.

The aims of this analysis are multifold. Firstly, we aim to carry out a direct comparison between the ages we estimate  and results obtained by  \citet{eta}  for a different metallicity range (hereafter Case 1, C1). Secondly, we aim to obtain a realistic estimate of the uncertainties in asteroseismic field halo star age estimates (C2). Finally, we aim to obtain an estimate of the uncertainty in the asteroseismic derived ages for GC stars (C3). 
Therefore, the sketched analysis was repeated assuming different sets of observational uncertainties in both classical and asteroseismic parameters. 

\begin{table}
    \centering
        \caption{Observational errors adopted in the Monte Carlo experiments for the three investigated scenarios.}
    \begin{tabular}{lcccc}
    \hline
    Scenario & $T_{\rm eff}$ & [Fe/H] & $\Delta \nu$ & $\nu_{\rm max}$\\
             & (K)    &  (dex) &   &   \\
    \hline
    C1 & 100 & 0.10  & 2.5\%  & 5\%\\
    C2 & 100  & 0.05 & 0.6\%  & 1.7\%\\
    C3 & 150 & 0.03 & 1\% & 4\%\\
    \hline
    \end{tabular}
    \label{tab:errori}
\end{table}

More specifically, the observational errors adopted for the C1 case were identical to those presented by \citet{eta}, who presented the same analysis for a population typical of the Milky Way disk;\ that is 2.5\% in $\Delta \nu$, 5\% in  $\nu_{\rm max}$, 100 K in $\rm T_{\rm eff}$, and 0.1 dex in [Fe/H]. The adoption of the same set of observational errors allows us to best disentangle possible differences and trends due to the differences in the stellar metallicity range covered by the two investigations. Moreover, these uncertainties are close to that reported by \citet{Stello2022} for the TESS analysis of RGB stars (i.e. 5\% in $\nu_{\rm max}$x and about 3\% in $\Delta \nu$), which provides useful insight into the uncertainty in age estimates achievable by TESS.

However, if one is interested in assessing the best precision achievable at present for asteroseismic age estimates for halo stars, we note that these observational errors are overestimated in a few cases. Recent works investigated asteroseismic data from the Kepler field \citep{Yu2018} to study halo stars \citep[e.g.][]{Montalban2021, Matsuno2021, Grunblatt2021}, reporting a median uncertainty in asteroseismic constraints of 0.6\% in $\Delta \nu$ and 1.7\% in  $\nu_{\rm max}$ and typical uncertainties in classical observables of 100 K in $T_{\rm eff}$ and 0.05 dex in [Fe/H]. These are the values adopted in C2.

Stars in GCs are an important part of the population of the halo.  Unfortunately, at present, asteroseismic observations for GC stars are still rare, with some data available for M4 GC. This situation is expected to improve in the future and some space missions have been proposed to explicitly target stars in clusters \citep[e.g. HAYDN,][]{Miglio2021}.
It is therefore interesting to compare asteroseismic age estimates for GCs with those obtained with the robust classic isochrone fitting method. Due to the relative high stellar density of GC stars and their large distances, observational uncertainties for such objects are higher than the mean uncertainties for field stars. In C3, we adopted an uncertainty for the asteroseismic observables of 1\% in $\Delta \nu$ and 4\% in $\nu_{\rm max}$ from \citet{Miglio2016}, while for classical data, we assumed an error of 150 K in $T_{\rm eff}$ and 0.03 dex in [Fe/H], which are typical values found in the recent literature. The error in $T_{\rm eff}$ comes from the adoption of scaling relations on photometric data, similarly to what was done in \citet{Malavolta14}.
The three investigated scenarios, and the associated uncertainties, are summarised in Table~\ref{tab:errori}.

\subsection{C1: Comparison of uncertainties in typical halo and disk stars}\label{sec:field_stars_c1}

\begin{figure*}
        \centering
        \includegraphics[height=5cm]{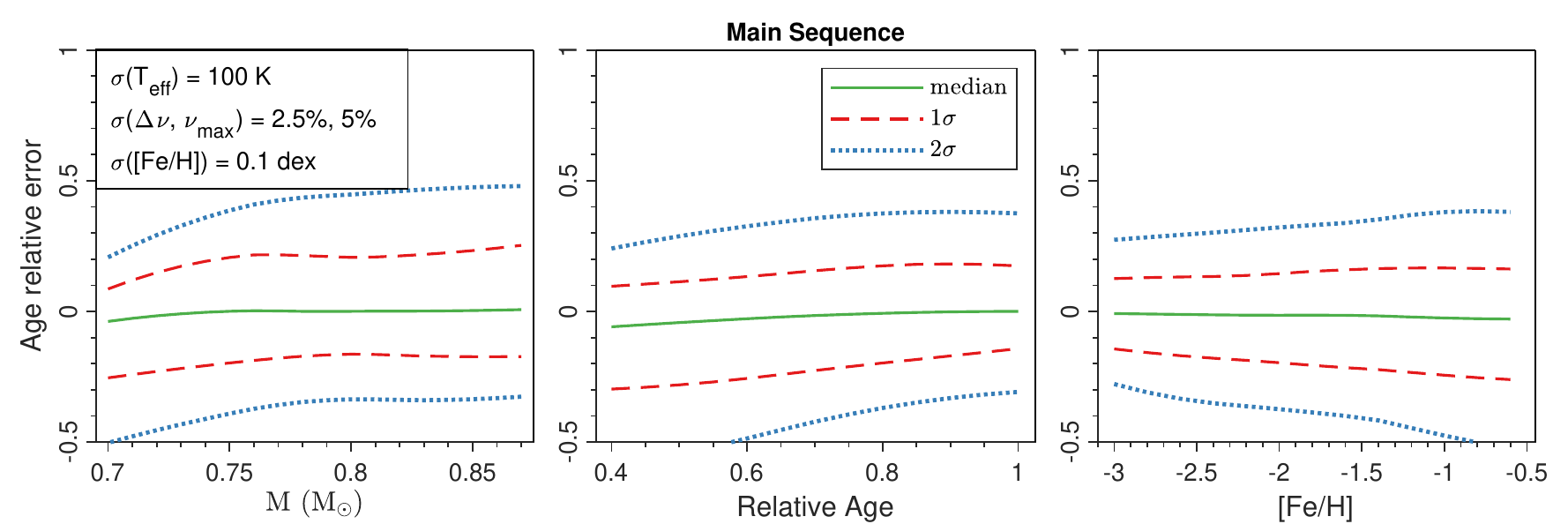}

        \includegraphics[height=5cm]{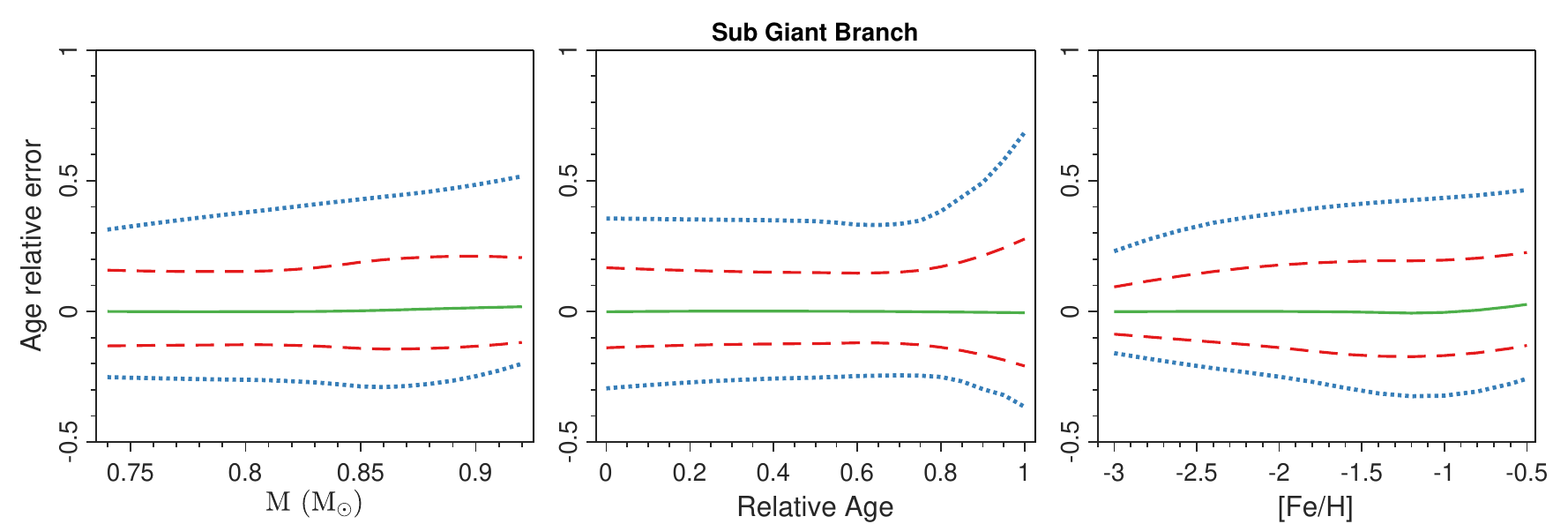}

        \includegraphics[height=5cm]{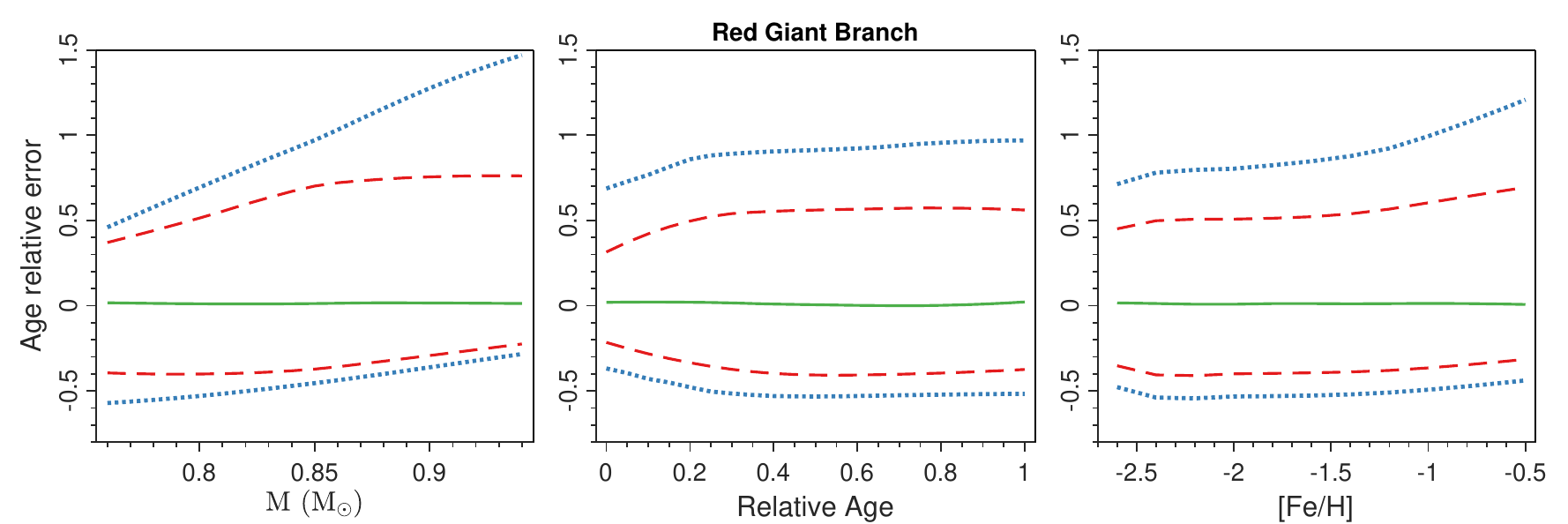}
        \caption{Relative errors on age estimates as a function of the true mass (left panels), 
                relative age (central panels), and metallicity [Fe/H] (right panels) of the
                star.
                From top to bottom, the panels show the relative error on age for MS, SGB, and RGB stars, respectively.
                The green solid line marks the error medians. The red solid line is the $1 \sigma$ error
                envelope, while the blue dashed one marks the position of the $2 \sigma$
                envelope (see text). A positive relative error indicates that the reconstructed age of the star 
                is overestimated with respect to the true one.}
        \label{fig:standard}
\end{figure*}

 Adopting the SCEPtER algorithm  (Sect.~\ref{Scepter_explain}), we estimated ages of synthetic stars $A_e$ and compared the results to actual values $A_t$ from the dataset.
Then the  relative age error is computed as
\begin{equation}
{\rm relative \;  age \; error} = \frac{A_e-A_t}{A_t},
\end{equation}
 and therefore a positive relative error indicates that the age of the star was
overestimated by the recovery procedure. 

Figure~\ref{fig:standard} shows the trend of the relative age errors versus
the true mass of the star, its relative age ---conventionally set to
zero at the start of the selected evolutionary phase and defined as the ratio between the current age of the star and its age at the end of the same evolutionary phase--- and its metallicity [Fe/H]\footnote{This is the present surface [Fe/H] value, which is different from the initial one due to microscopic diffusion.}.
 The adoption of relative age in the chosen evolutionary phase was preferred to global relative age at track level because the former allows a better synchronisation of the evolutionary stages after the central hydrogen exhaustion.

The figure also shows the relative error envelopes obtained by evaluating
the 16th and 84th quantiles (1 $\sigma$) and 2.5th and 97.5th quantiles (2
$\sigma$) of the  relative age error over a moving window\footnote{The
half width of the window is typically 1/12-1/16 of the range spanned by the independent variable. This choice allows us to maintain 
the mean relative error in the 1 $\sigma$ envelope owing to Monte Carlo sampling at a level of about 5\%, without introducing excessive smoothing.}.
The position of the 1 $\sigma$ envelope and of the median of the relative age error as a function of the true mass of the star and of its relative age are
reported in Appendix in Tables~\ref{tab:global_MS} to \ref{tab:global-pcage_RGB} in the section labelled C1.

The median relative errors are compatible with zero in all analysed cases, meaning that the recovery procedure is not intrinsically biased, as already verified for higher metallicities in \citet{eta}.
A $-5\%$ departure from zero is seen for MS stars with relative ages of lower than 0.5 due to an edge effect.
As the sample age range is fixed, lower relative ages correspond to lower masses,  which due to their long MS lifetimes are still far away from the TO.
Masses smaller than $M \sim 0.74$ correspond to the lower edge of the grid, and therefore their mass can only be overestimated, and the age underestimated.
For this reason, the distribution of errors for these lower masses is slightly shifted and the median falls slightly below zero, which could be falsely interpreted as a bias in the age estimation.
The edge effect can also be seen in the top-middle and top-left panels of Fig.~\ref{fig:standard}.
Edge effects were already pointed out in previous works \citep[e.g.][]{scepter1}, and therefore the estimation grid was purposefully designed to be larger with respect to the values of interest for this work in order to mitigate this problem.
The mass range of interest is about [0.8;  0.95] $M_{\sun}$, where the majority of observations of asteroseismic parameters for halo MS stars are expected to take place in the near future.

\begin{figure}
	\centering
	\includegraphics[width=9cm]{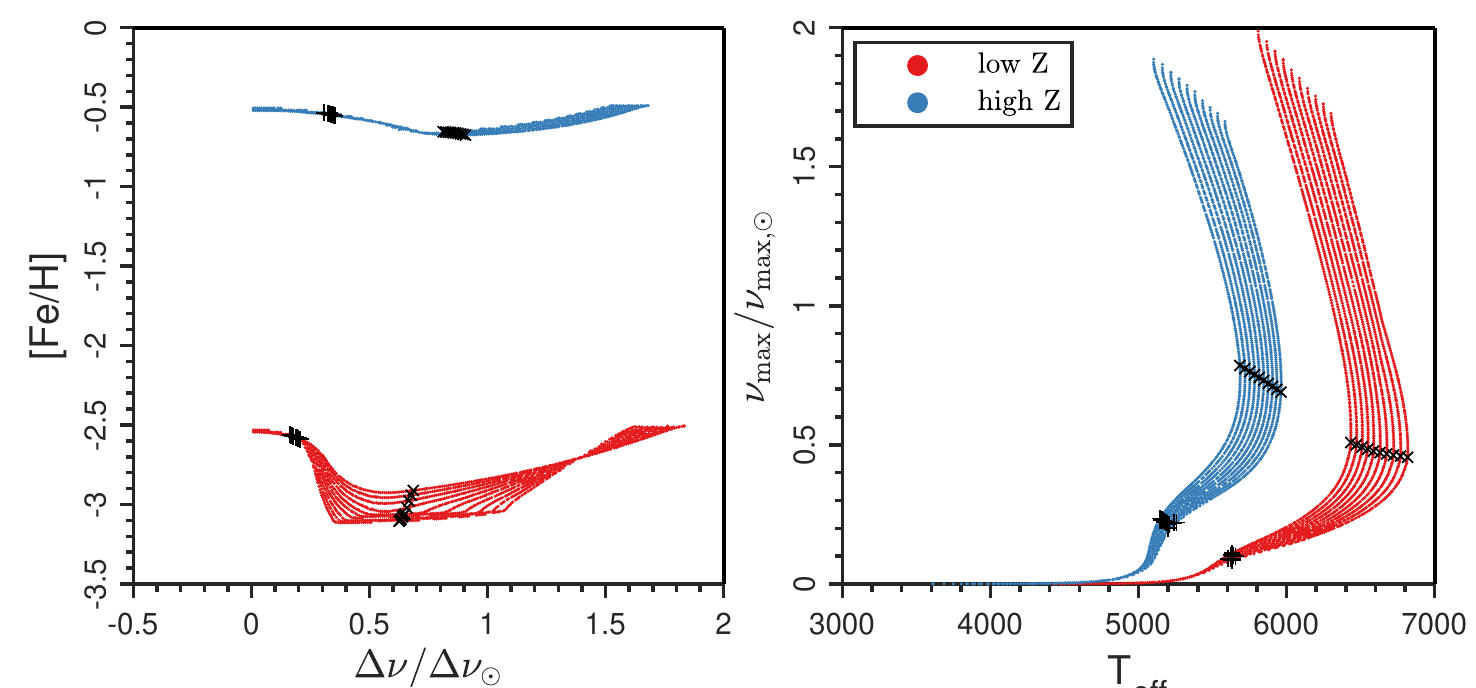}
	\caption{Distribution of identical stellar models ($M < 0.8$ $M_{\sun}$) at high ([Fe/H]$=-0.5$) and low ([Fe/H]$=-2.5$) initial metallicity. The black crosses "x" indicate the change in evolutionary phase from MS to SGB, while the black "+" signal the shift to RGB. Low metallicity models occupy a larger region in the observational parameter space.}
	\label{fig:confronto_grid}
\end{figure}

Apart from the mentioned edge effect, the relative errors on
ages for MS stars are almost constant, between 10\% and 20\%, as a function of mass and relative age.
Furthermore, the relative error on
ages slightly increases with metallicity. This effect arises from the grid morphology in the 4D observational parameter space. Lower metallicity models are more spread apart; therefore, given a set of observable uncertainties, it is easier to discriminate between solutions to find the correct one (see Fig. \ref{fig:confronto_grid}).
Typical relative errors on
ages obtained in this work are generally smaller than those reported by \citet{eta} for stars with chemical compositions typical of the Milky Way disk, by as much as 10\%, possibly due to an intrinsic difference in age estimation precision at different metallicity.
However, this comparison is not straightforward because 
the considered stellar population differs in terms of the mass range covered, with the stars sampled by  \citet{eta}  being more massive. As pointed out by \citet{eta}, the difference in the age estimation error can be as large as 15\% for stars of between 0.8 $M_{\sun}$ and  1.4 $M_{\sun}$ .
Furthermore, the  \citet{eta}  sample contains young MS stars, for which the relative errors on
ages can be larger than 100\%.

The relative errors on
ages for stars on the SGB do not change with the mass of the star and are generally constant throughout their entire, short lifetime, except for the final part where, as shown the central panel of Fig.~\ref{fig:standard}, the $1 \sigma$ and $2 \sigma$ error envelopes slightly increase.
During this final part of the SGB, theoretical models converge towards small observational parameters space because they are approaching the RGB, where the distribution of effective temperatures is narrower. The same happens to asteroseismic parameters, as they depend on $T_{\rm eff}$.
The packing of theoretical models makes the estimation process intrinsically more difficult, leading to larger age uncertainties.
The results again show an increase in age errors for higher metallicities, although in this case there are no previous studies for metallicity ranges for comparison.

Our analysis for RGB stars shows an important degradation in age estimation precision, with errors as large as 60\%.
The main reason for this degradation is again the packing of the models in the observable parameter space. 
As already reported in the literature, the models in this evolutionary phase have a narrow distribution in effective temperature, which is also much more dependent on the metallicity with respect to MS stars, and therefore the mass recovery is more difficult, leading to larger errors on  age \citep[see e.g.][]{Basu2010, bulge}.
Because of these large errors and the fact that negative relative errors cannot be greater than -100\%, the resulting age  distributions are slightly skewed towards positive errors.
Indeed, for more favourable conditions, in which age errors are smaller even for the RGB phase, distributions are again symmetric, as can be seen in the cases discussed below.

\subsection{C2: Effect of realistic uncertainties in halo field stars}\label{sec:field_stars_c2}

\begin{figure*}
\centering
\includegraphics[height=5cm]{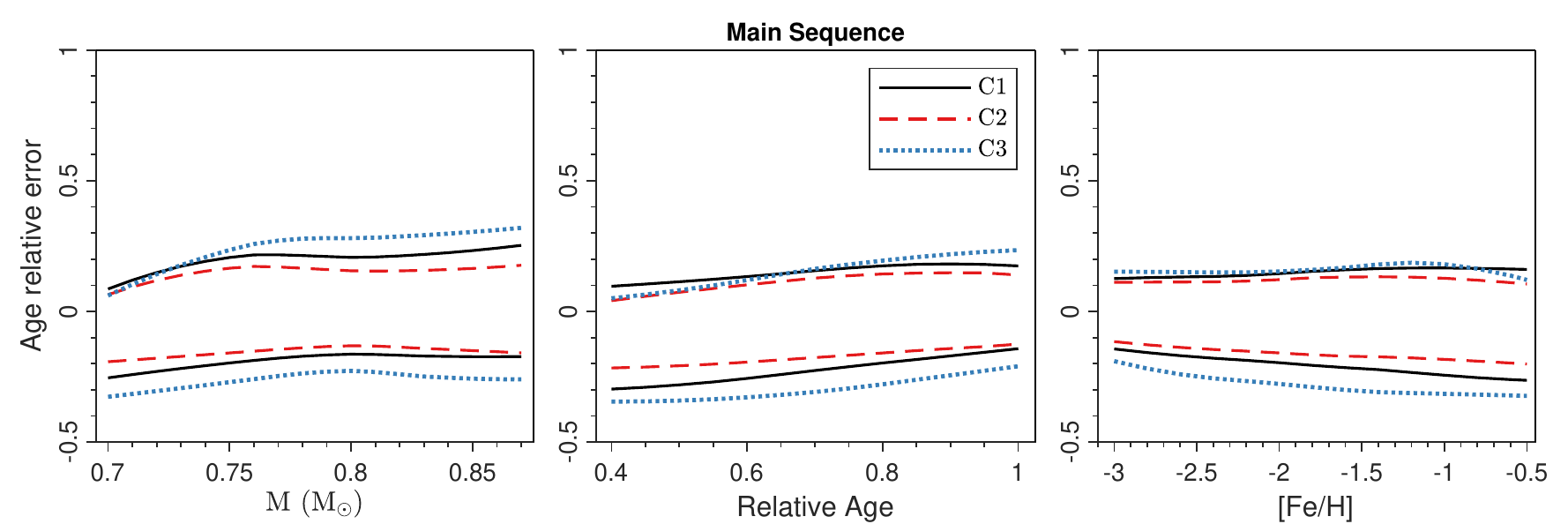}

\includegraphics[height=5cm]{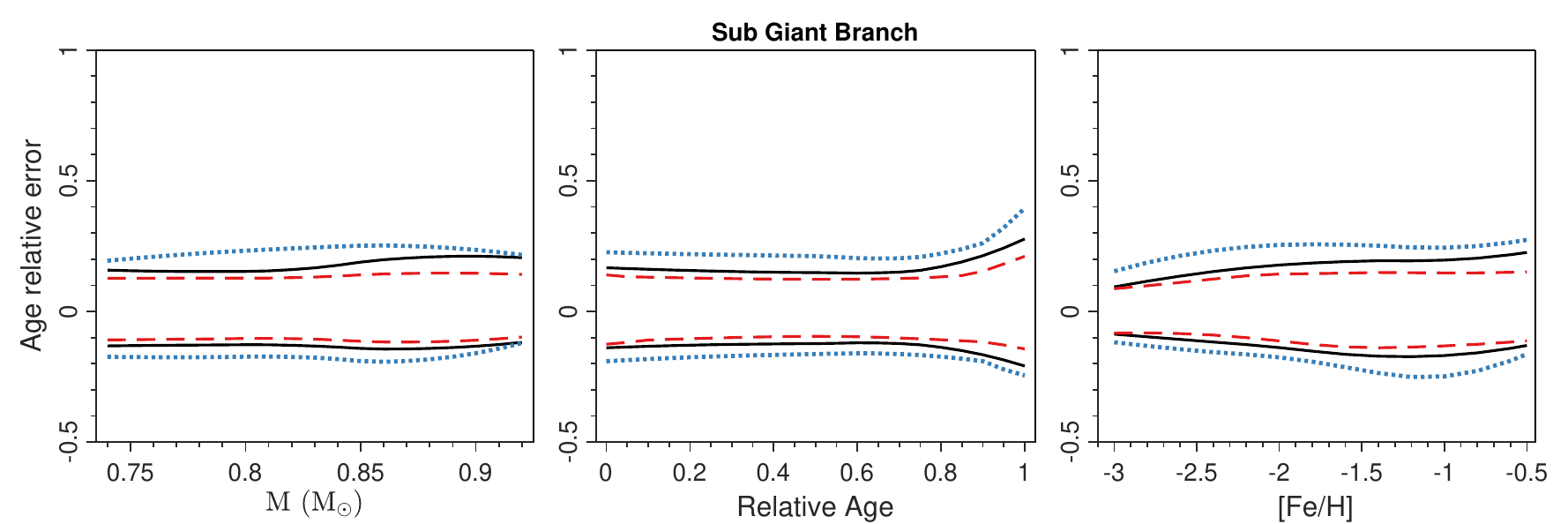}

\includegraphics[height=5cm]{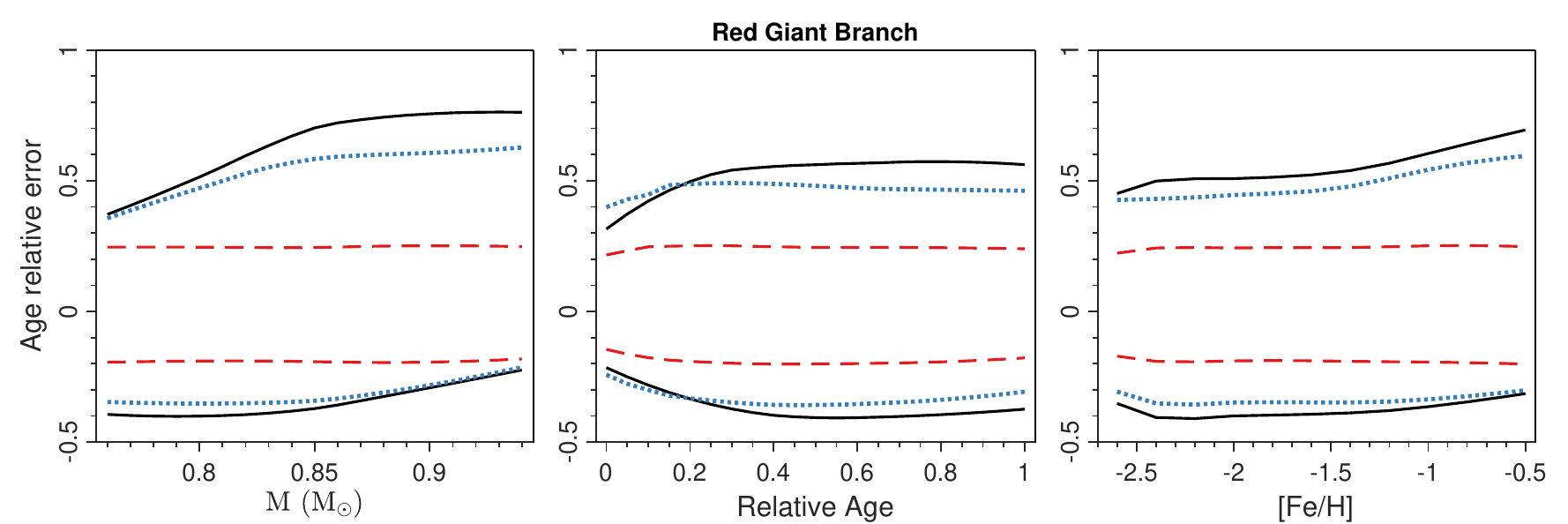}
\caption{Relative errors on age estimates in different evolutionary phases. {\it Top row}:  Relative
errors on age estimates as a function of true mass ({\it left panel}), 
                relative age ({\it central panel}), and metallicity [Fe/H] ({\it right panel}) for MS
                stars. 
                {\it Middle row}: Same as in the {\it top row} but for SGB stars. 
                {\it Bottom row}: Same as in the {\it top row} but for RGB stars. 
                Black lines mark the 1 $\sigma$ age estimation error envelopes computed adopting C1 uncertainties.
                The dashed red lines show 1 $\sigma$ envelopes for the C2 scenario.   
                The blue dotted lines mark 1 $\sigma$ envelopes for the C3 scenario (see text and Table~\ref{tab:errori}).}
\label{fig:Realscen}
\end{figure*}

The results of the analysis for C2 are reported in Tables \ref{tab:global_MS} to \ref{tab:global-pcage_RGB} under section C2 and are also shown in Fig.~\ref{fig:Realscen}, where 1 $\sigma$ envelopes are compared with those for the C1 and C3 scenarios.
During the MS and SGB phases, we note a reduction of $\sim 5$\% in the 1 $\sigma$ error envelopes, mainly due to the slightly better precision of $\rm T_{\rm eff}$.
Our finding that the  uncertainty on effective temperature is the main contributor to the  error on age estimations for MS stars is in agreement with the findings of \citet{Valle2018} for stars in the Milky Way disk.
Effective temperature is strongly linked to the mass of the star, and therefore a better constraint on $T_{\rm eff}$ leads to a better mass estimate, which in turn is highly correlated to stellar age.
During the MS phase, an improvement in asteroseismic data or in metallicity precision does not lead to a substantial decrease in the age estimation error.
For SGB stars, again, improving the precision in asteroseismic data does not have a significant effect. However, improving metallicity precision indeed leads to a significant reduction in the relative error on
ages, except for relative ages above 0.8, where models occupy a smaller region in the observational parameter space.

For RGB stars, the reduction in errors  on age estimations is significant, namely as large as 20\%--30\%.
The smaller uncertainties on age estimations lead to symmetric error envelopes, confirming that the asymmetry seen in the standard case was due to large observational errors.
The largest contribution to the error reduction is the improvement of the asteroseismic data, again in agreement with the results of \citet{Valle2018} for more metal-rich stars.
Reducing the error in effective temperature does not have a noticeable effect.
The evolutionary tracks of RGB models are very close in effective temperature, and therefore an unrealistically small uncertainty in effective temperature would be needed to efficiently improve the recovery of the stellar mass and age.
Indeed, with an error of $\sim 80$ K on $T_{\rm eff}$ , a 3 $\sigma$ confidence interval covers almost the entire range of possible effective temperatures for RGB stars.
Finally, the improvement of the metallicity estimate has almost no effect on the precision of age recovery.

\subsection{C3: Effect of realistic uncertainties for GC stars}\label{sec:GC}

The results for the C3 scenario are shown in Tables~\ref{tab:global_MS} to \ref{tab:global-pcage_RGB} in the corresponding sections.
Figure~\ref{fig:Realscen} shows the 1 $\sigma$ error envelopes as a function of true mass, relative age, and metallicity compared with the results for the standard and field stars cases.
In MS and SGB phases, the typical error is slightly larger than in the standard case.
This is due to the larger uncertainty on $T_{\rm eff}$, although this error increase is partially counterbalanced by the reduction in the uncertainty of the other observables, especially of the metallicity.

However, in the RGB phase, the slight reduction in the asteroseismic data error leads to a noticeable reduction in the age relative uncertainty.
Therefore, for RGB stars in GCs (the only ones with available observational data in addition to horizontal branch stars, which are not studied in this work), one expects a typical uncertainty on age estimates of about 35\%--45\%.

\section{Stellar model uncertainty propagation}
\label{sec:errorprop}

When grid-based techniques are applied to real stars rather than 
to synthetic ones, the accuracy of age estimates depends on the 
systematic discrepancies between the adopted stellar models and observational data.
The uncertainties on the input physics and physical mechanisms adopted in stellar evolutionary codes leads to indeterminacy in the grid-based results.
This issue has been extensively discussed in many works \citep[e.g.][]{Basu2012, Lebreton2014, scepter1, eta}, and it is well known that the variability in age estimates obtained by different estimation pipelines is of the same order as the statistical uncertainty on the age as obtained by a single pipeline \citep{Basu2012, eta, SilvaAguirre2017, smallsep, Tayar2022}. The aim of this section is to quantitatively analyse some of the bias occurring as a result of the different modelling choices  for age estimates of old, metal-poor stars.

We focus our analysis on the uncertainty in the mixing-length parameter and in the efficiency of element diffusion. Contrary to the work of \citet{eta}, we do not take into account the uncertainty in the original helium abundance. This is because \citet{eta} showed that the effect of this uncertainty is negligible; moreover for low-metallicity stars, the variations in the initial He abundance due to different $\Delta Y/ \Delta Z$ values is not significant.

We performed these estimates following the approach described by \citet{scepter1}. More specifically, we computed some non-standard grids of perturbed stellar models by individually varying the chosen inputs to their extreme values, while keeping all the other inputs fixed to their reference values.
Artificial stars were then sampled from these grids, and their ages were estimated using the standard grid.
In both cases, the results of which are presented in Sect.~\ref{sec:ml} and Sect.~\ref{sec:feh}, the assumed uncertainties on the observables are those of the C2 case: $\sigma(T_{\rm eff}) = 83$ K, $\sigma(\Delta\nu) = 0.6$\%, $\sigma(\nu_{\rm max}) = 1.7$\%, and $\sigma(\rm [Fe/H]) = 0.05$ dex.

\begin{figure*}
\centering
\includegraphics[height=5cm]{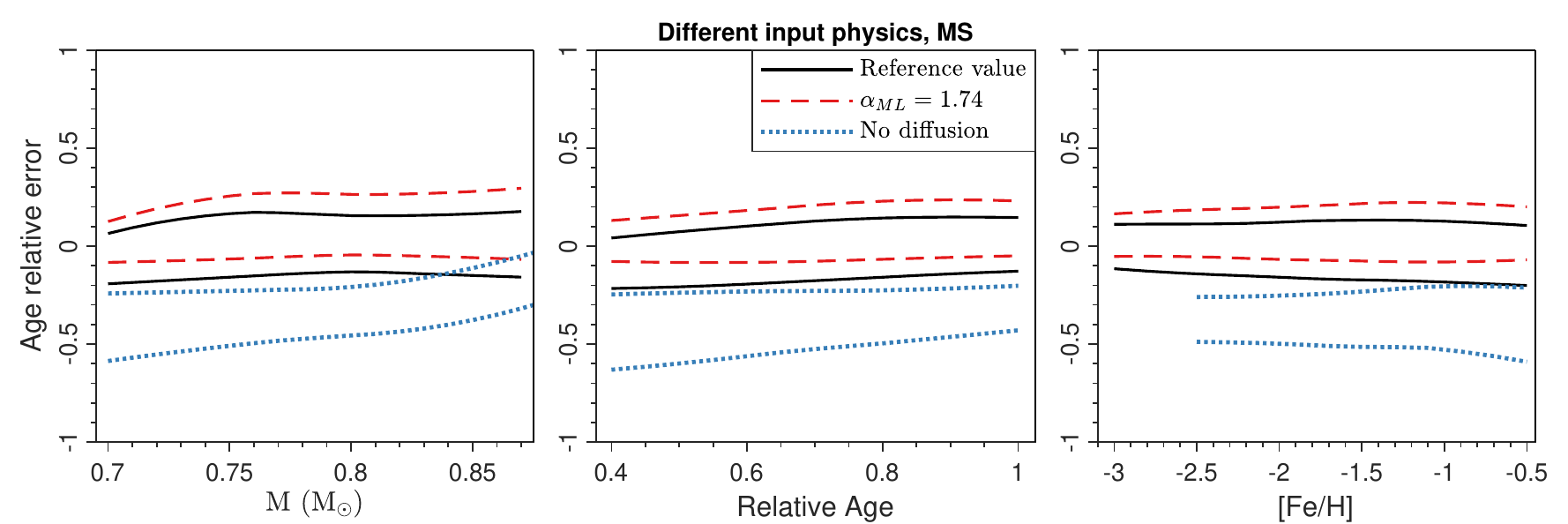}

\includegraphics[height=5cm]{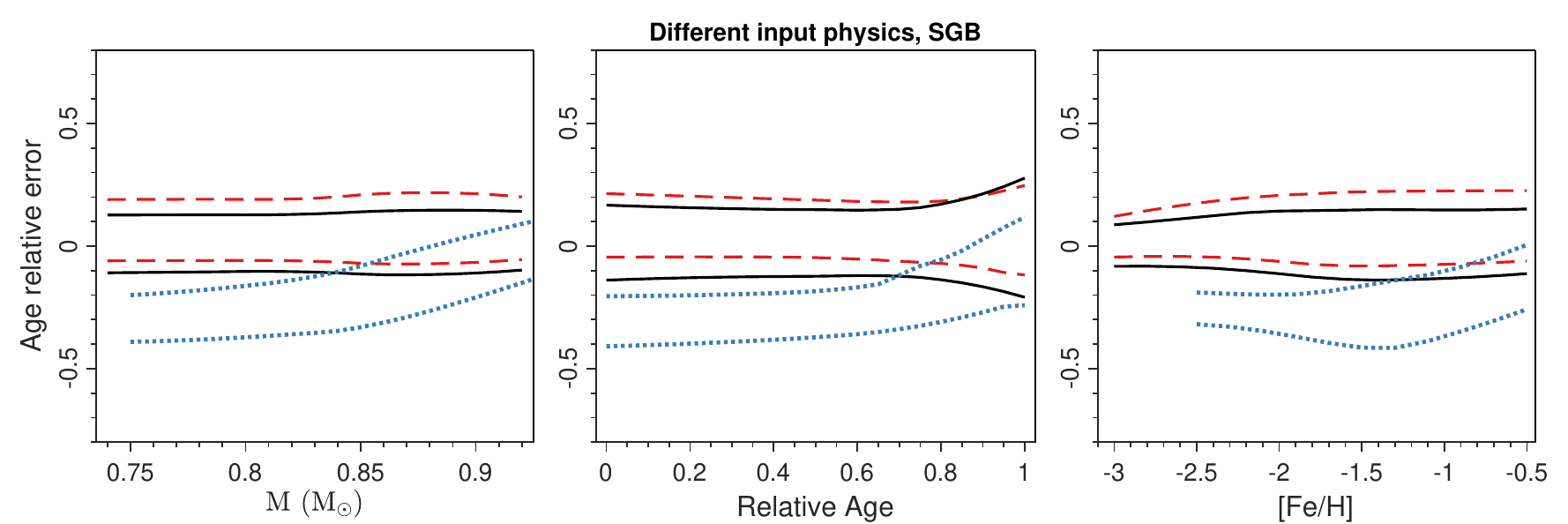}

\includegraphics[height=5cm]{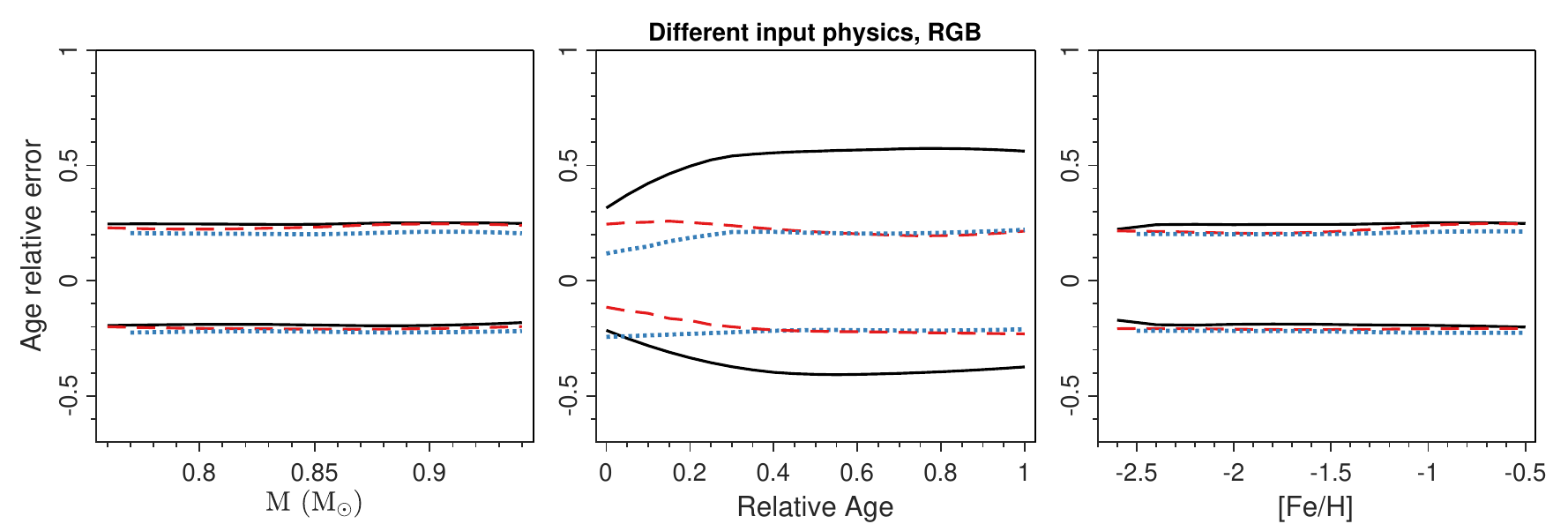}
\caption{Envelope of relative errors on age estimates  as a function of mass
  and relative age of the stars, with different values for the input physics compared to the estimation grid. The black line corresponds to the standard case;
  the red one to a different efficiency of the external convection; the blue one to stars without diffusion whose age is estimated by adopting a grid of models with diffusion.} 
\label{fig:input}
\end{figure*}

\subsection{Mixing-length value} 
\label{sec:ml}

It is increasingly apparent that the use of a solar-calibrated mixing-length
value for stars that differ from the Sun in mass, composition, and/or 
evolutionary phase may not be appropriate \citep[see e.g.][]{Deheuvels2011, Bonaca2012, Mathur2012,  Tanner2014, Trampedach2011, Magic2014, Yildiz2007, Clausen2009}.
To quantify the effect of varying the efficiency of the super-adiabatic convective transport, we computed one additional non-standard grid of stellar models by assuming the solar calibrated mixing-length parameter $\alpha_{\rm ml}$ = 1.74. We then built a synthetic dataset of $N = 30\,000$ artificial stars (for the evolutionary phases discussed above), sampling them from the non-standard grid.
The ages of the synthetic stars were then estimated using the standard grid for the recovery, which assumes our M4 calibrated value of $\alpha_{\rm ml}$ = 1.90.
The results of these tests are presented in section `$\alpha_{\rm ml} = 1.74$' in the tables presented below.

The bias for MS stars is noticeable, with a departure of the median from zero of about 10\%, meaning that, in general, the ages of the old stars in the halo can be overestimated by as much as 1 Gyr if the efficiency of convection is not well parameterized.
Artificial stars with varied $\alpha_{\rm ml}$ occupy a different location 
in the 4D space of the observable quantities with respect to standard models 
of the same mass and age.
In particular, models with a less efficient convection (smaller $\alpha_{\rm ml}$) are colder than their corresponding model with $\alpha_{\rm ml} = 1.90$). In addition to this direct effect, which is the main factor responsible for the age bias \citep[see the analysis in][]{Valle2018}, the mixing length value modification also has an effect on the asteroseismic data, which depend on $\rm T_{\rm eff}$ through the scaling relations of Eqs.~(\ref{eq:dni}) and (\ref{eq:nimax}).
This difference in the observable parameters leads the algorithm to incorrectly select stars with smaller masses that, because of the longer evolutionary times, are usually older, which explains the positive age bias.

During the SGB phase, the distance in the parameter space among evolutionary models with different mixing-length efficiency is reduced compared to the MS case; however, the effect is still present, resulting in a smaller but still present bias that is no larger than 5\%.
In the RGB phase, there is a trend inversion.
At the start of the RGB, the bias is still small and positive in accordance with what can be seen at the end of the SGB phase. However, for increasing values of relative age, the bias reduces, until it reaches zero at around 30\% of the RGB lifetime.
The bias in the age estimation is then inverted, and for more evolved RGB models, the algorithm incorrectly selects models with a higher mass, meaning generally younger stars, leading to a small negative bias that does not exceed 4\%.

This trend is caused by the progressive clumping of the grid in the observational parameter space, which becomes more severe as the evolution on the RGB proceeds.
In particular, during RGB evolution, the difference in effective temperature among models of different mass drastically reduces. This effect, coupled with the relatively large uncertainty in $T_{\rm eff}$, means that, contrary to the case for stars on the MS, the algorithm cannot use this observable to efficiently recover mass and age.
Instead, the stellar age recovery is mainly based on the asteroseismic parameters, as discussed in \citet{Valle2018}. 
It is possible to show that, at fixed mass, metallicity, and evolutionary phase, the asteroseismic parameters are smaller for a lower efficiency of the convection. This means that the algorithm is lead to choose a more massive model, which causes the negative age bias. The fundamental and different role of the effective temperature was checked by direct computation, repeating the age estimation without  the effective temperature observational constraint.  While the results did not change in the RGB phase, the age bias in the MS reversed, becoming negative, thus confirming the dominant impact of $T_{\rm eff}$ in this phase. The effects of the biases in effective temperature and asteroseismic parameters are of the same order but act in opposite directions, thus leading to a partial cancellation.
Ultimately, this result is further evidence  supporting the finding that grid-based estimates and their biases are extremely dependent on the adopted observational constraints  and on the studied evolutionary phases \citep[see e.g][]{Basu2012, binary}.  

In conclusion, it is interesting to note that the bias for low-metallicity stars is generally lower than that for the typical disk stars evaluated by \cite{eta} (the comparison stands only for MS stars as the other two evolutionary phases were not studied).
This trend in metallicity is also visible in the top-right panel of Fig.~\ref{fig:input}.
This is likely a result of the grid being more compact for less metallic stars, as already discussed. This leads to models with different convection efficiency becoming closer in the parameter space, in particular in the asteroseismic parameters, which means biases get smaller.

\subsection{Element diffusion} 
\label{sec:feh}

\citet{scepter1} discussed the importance of considering the effects of microscopic diffusion when determining stellar parameters by means of grid-based techniques, assessing the bias in mass and radius estimates when element diffusion is neglected.
While microscopic diffusion has been proved to be efficient in the Sun \citep[see e.g.][]{Bahcall2001}, with a related 15\% uncertainty, its efficiency in Galactic GC stars is still debated \citep[see e.g.][]{Gruyters2014, Nordlander2012, Gratton2011, Korn2007}.

To conservatively estimate the possible age bias due to microscopic diffusion effects, we sampled stars from a non-standard grid with models computed without diffusion, and then recovered their age through the standard grid, where diffusion is taken into account.
Results are shown in Fig.~\ref{fig:input} and tabulated in the sections labelled `no diffusion' in all of the tables for different evolutionary phases below.
As already seen in previous works \citep{scepter1, eta}, biases can be relatively large for stars on the MS, with values of around 40\%.The effects increase with decreasing stellar mass, because the slower MS evolution  allows more time for diffusion to change the chemical profile of the structure. 
This large bias arises from both the evolutionary time change and the surface temperature and chemical composition variation due to diffusion.
In general, the mass recovered is biased towards higher masses, leading to large negative bias for the estimated ages.

However, totally neglecting diffusion is probably an overly crude assumption; in this way we conservatively obtained an upper bound for the possible age bias. A realistic bias due to microscopic diffusion uncertainty would be smaller, though still important.

A small bias persists during the major part of the SGB evolution, at least until the first dredge-up negates most of the diffusion effects on the chemical composition.
Finally, during the RGB phase, these biases are negligible because of the completion of the first dredge-up and the RGB evolutionary timescale, which is much shorter than the diffusion timescale. As discussed above, a more realistic scenario would lead to even smaller biases.

\section{Asteroseismic and classical age of the globular cluster M4}\label{sec:M4}

In this section, we apply the asteroseismic age estimation technique to some stars of the GC M4, for which asteroseismic parameters are available in the literature. The estimated mean age  is then compared to that obtained adopting the classical isochrone-fitting method. The analysis is similar to that presented  by \citet{Miglio2016}.  \citet{Miglio2016} presented global asteroseismic parameters from the K2 mission campaign, which detected clear solar-like oscillations for seven RGB stars and one star in the red part of the horizontal branch (HB).

The present analysis includes the seven RGB stars only. The HB star investigation is avoided because  this evolutionary phase is not included in our
grid of models; moreover the positions of the stars inside the HB region depend on the stochastic mass loss during the RGB phase, and so modelling HB stars would introduce other sources of systematic uncertainty.

For the analysis, we adopted the average metallicity of the cluster available in the literature, namely [Fe/H] = -1.17 $\pm$ 0.03 \citep{Bailin2019}.
Effective temperatures adopted in \citet{Miglio2016} came from \citet{Marino2008}, derived from spectroscopic data.
However, a direct comparison showed that these values do not agree well with the theoretical isochrone that better fits the colour--magnitude diagram of the cluster, 
a problem also evident in the paper by \citet{Miglio2016}.
Therefore, we derived new values for the effective temperature, adopting a widely used metallicity-dependent relation between stellar colour and temperature:
\begin{equation}
    \theta = b_0 + b_1 C + b_2 C^2 + b_3 {\rm [Fe/H]} + b_4 {\rm [Fe/H]}^ 2+b_5 {\rm [Fe/H]} C,
    \label{CtoT}
\end{equation}
where $T_{\rm eff}=5040/\theta$ and $C$ is the colour of the star in a given photometric system.

We used seven colours for each star: four colours from the \citet{Stetson2019} catalogue  (classical photometric bands from the Johnson-Cousins filter system) and three colours from the third early data release of Gaia \citet{EDR3phot}.
Coefficients $b_i$ for the Johnson-Cousins colours can be found tabulated in \citet{Ramirez2005}, while the more recent coefficients for the Gaia colours were obtained from \citet{Mucciarelli2020}.
For each star, the temperatures derived from the seven colours were averaged to compute a single estimate of the effective temperature, while errors were computed through error-propagation rules using uncertainties given in these latter respective works.
No correction due to differential reddening was adopted, as our results are very close to the ones presented in \citet{Malavolta14}, where such a correction was indeed made.
Table~\ref{tab:M4data} summarises the data; it can be seen that our estimates of effective temperature are generally 100 K higher than those in \citet{Marino2008}; however, they are in good agreement with the values from \citet{Malavolta14}, which were also obtained from photometric data (although their dataset does not cover the entire sample of stars for which asteroseismic parameters are available).
The computed values are in better agreement with extrapolated temperatures from the theoretical isochrone that fitted the cluster CMD.

\begin{table}[!]
\caption{Physical properties for the seven RGB stars for which \citet{Miglio2016} reported unequivocal solar-like oscillations.}
\centering
\small
 \begin{tabular}{ c | c | c | c | c } 
 \hline
ID      &$\Delta\nu$    &$\nu_{\rm max}$    &$T_{\rm eff}$ [K]            &$T_{\rm eff}$ [K]  \\
        &[$\mu$Hz]     &[$\mu$Hz]     &$\pm$ 100 K               &$\pm$ 150 K \\ 
        &               &               &\citet{Miglio2016}        &This work    \\ 
        \hline
S1      &1.83$\pm$0.02  &11.1$\pm$0.4   &4585                     &4671\\
S2      &2.55$\pm$0.04  &17.2$\pm$0.7   &4715                     &4767\\
S3      &2.62$\pm$0.04  &17.7$\pm$0.7   &4710                     &4815\\
S4      &2.64$\pm$0.02  &18.5$\pm$0.7   &4715                     &4801\\
S5      &4.14$\pm$0.02  &32.5$\pm$1.3   &4847                     &4964\\
S6      &4.30$\pm$0.02  &32.9$\pm$1.3   &4842                     &4976\\
S7      &4.30$\pm$0.02  &34.3$\pm$1.4   &4805                     &4915\\
 \hline
 \end{tabular}
 \tablefoot{The first column reports the ID used in \citet{Miglio2016} to identify the seven RGB stars. The second, third, and fourth columns show the global seismic parameters and the effective temperatures from \citet{Miglio2016}.   The effective temperatures computed in this work are shown in the last column.}
\label{tab:M4data}
\end{table}

\subsection{Stellar models grid}

The estimation grid was again computed using the FRANEC code.
The [Fe/H] range was chosen to be [-1.3; -0.9] in order to comfortably contain the 3 $\sigma$ confidence interval for the estimated value of [Fe/H]=-1.17. The grid includes a total of 16 different metallicities with finer metallicity steps close to the central value. The value for metallicity, $Z$, was derived from iron abundance, assuming an $\alpha$-enhancement of 0.4 \citep{Marino2008}.

\begin{figure}
        \centering
        \includegraphics[height=10cm]{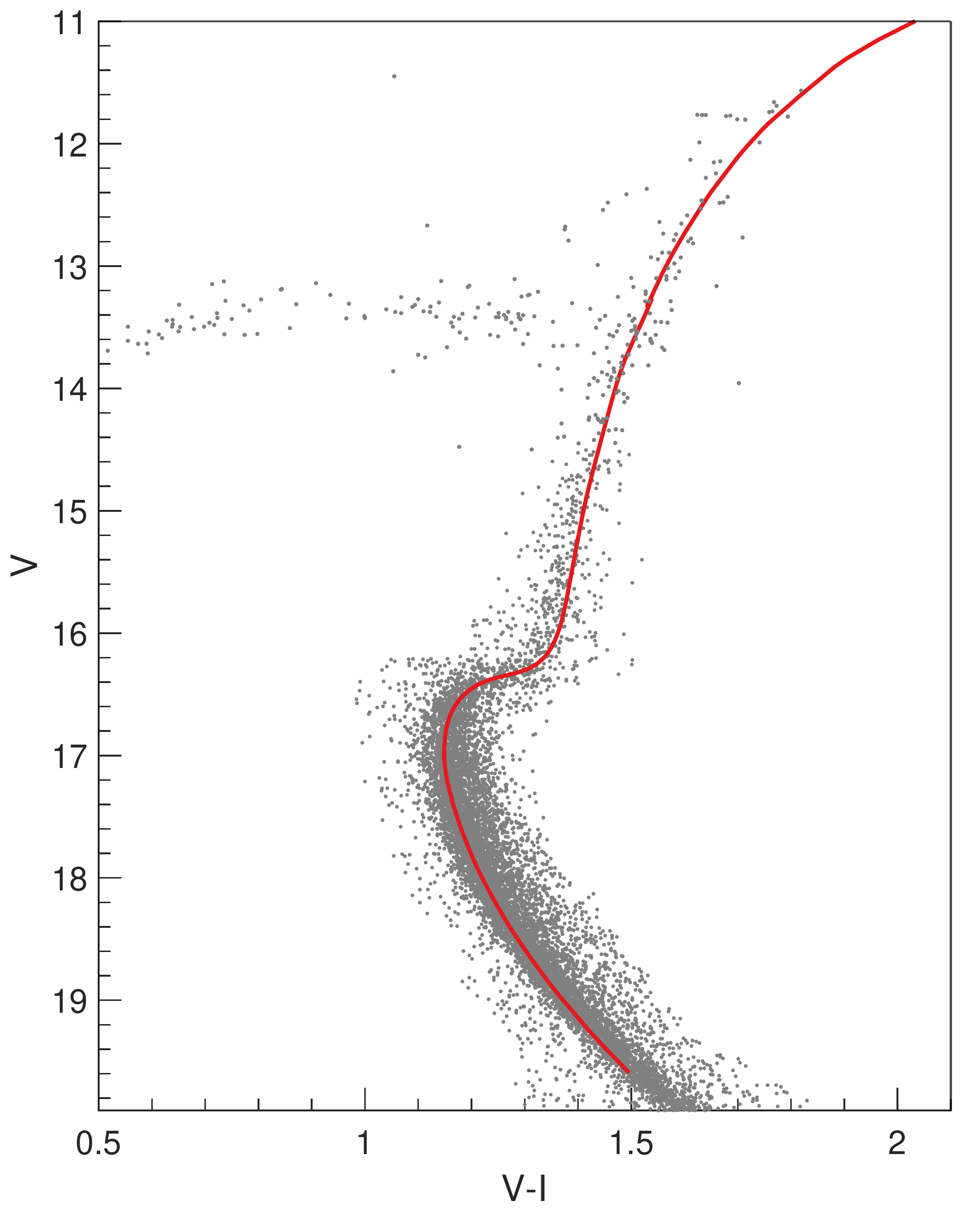}
        \caption{Theoretical 12 Gyr isochrone red line superimposed to M4 data from \citet{Stetson2019}. The isochrone was computed for [Fe/H] = -1.17 and $Y$ = 0.252. The value of the mixing-length parameter $\alpha_{\rm ml} = 1.90$ was chosen to best represent the RGB region. The adopted distance modulus and reddening are from \citet{Hendricks2012}.}
        \label{fig:Iso12}
\end{figure}

The adopted mass range is [0.76; 1.10] $M_{\sun}$, which was obtained from a trial-and-error procedure and is large enough to keep the mass estimation results far from the edges of the grid, thus helping to avoid the  previously discussed edge effects.
The value of the mixing-length parameter $\alpha_{\rm ml}$ was set to 1.90, because theoretical isochrones computed adopting this value were seen to agree well with the photometric RGB of M4 (see Fig.~\ref{fig:Iso12}).

As the evolutionary phase of the stars is known, we only kept models in the RGB in the grid. This implied removing stars with a mass smaller than $\sim 0.80$ $M_{\sun}$ as they have not yet reached the RGB.
From the literature it is known that the M4  TO mass is around 0.84 $M_{\sun}$ which, coupled with  poor precision in age and mass estimation for RGB stars mentioned above, means that results distributions were likely to undergo a small deformation due to edge effects.
Further edge effects are introduced because we eliminated all models older than the Universe from the estimation
grid, as they would lead to implausible results.From the literature, we expect M4 to have an age of about 12 Gyr, and with predicted errors for the asteroseismic method of the order of 40\%, the estimated age confidence intervals could overcome the present age of the Universe.
However, we quantified these edge effects simply by repeating the computation without cutting the grid and comparing the results. It was found that central values of age and mass differ by no more than 4\% and 1\%, respectively, between the two cases. Therefore, the decision to exclude models with implausible ages appears justified.

\subsection{Results}

\begin{table}[!]
\caption{Mass, radius, and age inferred using SCEPtER for the seven RGB stars observed in M4 from \citet{Miglio2016}. For each quantity is reported the central value, (q50) and the 1 $\sigma$ envelope (q16, q84).}
\centering
\small
 \begin{tabular}{ c | c | c | c } 
 \hline
quantile    &Mass           &Radius         &Age                \\
            &[$M_{\sun}$]  &[$R_{\sun}$]  &[Gyr]            \\
\hline
\multicolumn{4}{ c }{S1}\\
\hline
q16         &0.82           &15.94          &8.26                  \\   
q50         &0.85           &16.18          &12.35                  \\
q84         &0.94           &16.79          &13.50                   \\
\hline
\multicolumn{4}{ c }{S2}\\
\hline
q16         &0.82           &12.77          &7.15             \\
q50         &0.86           &13.03          &11.66            \\
q84         &0.98           &13.68          &13.43            \\
\hline
\multicolumn{4}{ c }{S3}\\
\hline
q16         &0.82           &12.54          &7.50            \\
q50         &0.85           &12.77          &12.08           \\
q84         &0.97           &13.37          &13.45           \\
\hline
\multicolumn{4}{ c }{S4}\\
\hline
q16         &0.84           &12.60          &5.61     \\
q50         &0.93           &13.06          &8.64    \\
q84         &1.06           &13.62          &12.73     \\
\hline
\multicolumn{4}{ c }{S5}\\
\hline
q16         &0.83           &9.32           &6.86                \\
q50         &0.89           &9.53           &10.05            \\
q84         &1.00           &9.93           &12.94    \\
\hline
\multicolumn{4}{ c }{S6}\\
\hline
q16         &0.82           &9.08           &9.73         \\
q50         &0.84           &9.13           &12.69          \\
q84         &0.90           &9.34           &13.35               \\
\hline
\multicolumn{4}{ c }{S7}\\
\hline
q16         &0.83           &9.09           &6.82        \\
q50         &0.89           &9.30           &10.28     \\
q84         &1.00           &9.68           &12.86        \\
\hline
\end{tabular}
\label{tab:results_m4}
\end{table}

Table~\ref{tab:results_m4} shows the results of mass, radius, and age estimation for the seven stars of M4.   The central values are reported as q50, while q16 and q84 represent the lower and upper 1 $\sigma$ confidence interval.
Due to the fast evolution of stars past the TO, it is possible to assume that the mass of RGB stars is close to the TO mass.
Computing the median of all the obtained q50 values, it is possible to estimate the TO mass, namely $M = 0.86 \pm 0.04$ $M_{\sun}$, where the uncertainty corresponds to a $1 \sigma$ level confidence interval and is derived considering both the variability between stars and the intrinsic statistic uncertainty in the fitted masses.
This result is in good agreement with the estimate by \citet{Miglio2016} and with the TO mass found from isochrone fitting, which is around 0.85 $M_{\sun}$.

\begin{table}
\caption{Age estimates for NGC 6121 found in the literature.}
\centering
 \begin{tabular}{ l  c } 
 \hline\hline
Reference           &Age [Gyr]          \\ \hline
\citet{Caputo_85}    &12$\pm$2           \\
\citet{Hansen_2002}  &12.7$\pm$0.7       \\
\citet{Salaris_02}   &11.7$\pm$1.1       \\
\citet{Dotter}       &12.5$\pm$0.5   \\
\citet{Vandenberg_2013}    &11.5$\pm$0.4  \\
\citet{Wagner_17}    &13.493$^{+0.007}_{-0.027}$\\
 \hline
 \end{tabular}
\label{tab:age}
\end{table}

In particular, it can be seen that the lower error in age is around 30\%, as reported in Table~\ref{tab:global_RGB} and Table~\ref{tab:global-pcage_RGB}, for stars with $M \sim 0.86$ $M_{\sun}$ and relative age $\sim 0.9,$ which is where the RGB bump occurs.
The median age, which we take as an estimate of the cluster age, is $11.9 \pm 1.5$ Gyr, where the uncertainty is obtained as described for the TO mass. This value is in good agreement with results from the literature \citep[e.g.][see Table~\ref{tab:age}]{Vandenberg2013, Dotter2010, salaris2002}, although errors are very large, as is expected given the analysis in Sect.~\ref{sec:GC}.
However, the upper limit on the error on stellar age  has been artificially reduced by removing all the models older than the Universe from the grid. Therefore, we see an upper error of $\sim 10\%$ instead of $\sim 40\%$ (see Fig.~\ref{fig:M4results} in Appendix).

The previous results were obtained after the star labelled S4 in \citet{Miglio2016} was removed from the analysis. It was deemed to be an outlier due to its estimated age and mass being significantly different from the other six stars.

In closing, it is worth discussing the fact that the cluster age was obtained without imposing a common age or composition in the fit of single stars. This approach confers the advantage of straightforward implementation \citep[e.g.][]{Gai2011, Sandquist2016}. However, as extensively discussed by \citet{cluster2018}, whenever common age and composition constraints are imposed, the  cluster-age estimation significantly shrinks towards the common value. This suggests that a different approach to cluster-age estimation ---for example by Markov Chain Monte Carlo (MCMC) methods--- could possibly be used to obtain a precision comparable to that achieved with classical methods. However, a detailed computation of the cluster age by this different technique is outside the scope of the present paper.  

\section{Conclusions}\label{sec:conclusions}

We performed a theoretical investigation to quantify the uncertainty on the age estimation of field and cluster stars with characteristics typical of the halo of the Milky Way.
To do this, we employed the SCEPtER pipeline to estimate the ages of a large sample of synthetic, metal-poor, and old stars in the MS, SGB, and RGB evolutionary phases in order to mimic halo field stars. We used both classical ($T_{\rm eff}$, [Fe/H]) and asteroseismic ($\Delta\nu$, $\nu_{\rm max}$) observables.

We investigated several scenarios characterised by different adopted observational uncertainties.
First, to highlight, in a consistent way, differences in age estimation precision between halo and disk populations, we adopted the same observational uncertainties as those used by \citet{eta} for disk stars. 
We found that the asteroseismic age estimates are more precise for old metal-poor stars as compared to more metallic stars. This is because the less metallic stars are usually more spread apart in the adopted observational constraint space; this makes it easier to reconstruct the model mass, which is tightly correlated with age.
We detect a large discrepancy in the  precision of age estimations between MS/SGB and RGB stars, with the former being recovered with a typical precision of around 10\%--20\%, while the latter can only be recovered with much lower precision ($\sim 60\%$). This discrepancy stems again from the morphology of the estimation grid, which is much more clumped for RGB stars.

However, at present it is only possible to achieve somewhat smaller uncertainties than the ones adopted in \citet{eta}. Therefore, we repeated the analysis by assuming typical uncertainties in the observables from the recent works by \citet{Montalban2021} and \citet{Matsuno2021}.
Results show a slightly better precision of age estimation for MS and SGB stars, which is mainly due to better constraints on effective temperature and metallicity. However, for RGB stars, the improvement is relevant: the average expected uncertainty decreases to about 20\%. This large error reduction is mainly due to the improvement of the observed asteroseismic parameters.

To analyse the effects of systematic discrepancies between grid models and observed stars we evaluated the effects of the still present uncertainties affecting the convection and microscopic diffusion efficiency  on age estimations, as already done in \citet{eta} for disk stars.
We find that a difference in the mixing-length parameter between synthetic stars and the estimation grid models leads to significant biases when estimating the ages of MS stars; this difference is caused by the effective temperature differences. These discrepancies cause the  algorithm to mistake the mass of the star and consequently its age.
This bias is smaller for SGB stars because models with different $\alpha_{\rm ml}$ are closer to one another in the observational parameter space compared to MS stars.
In the RGB phase, the bias is opposite. In this evolutionary phase, the discriminating power of the effective temperature is much reduced because of the packing of the tracks. In this situation, the fit is mainly based on the values of the asteroseismic parameters, which lead to a preference for more massive and therefore younger models.
This inversion confirms that generalizing-grid-based results for evolutionary phases or observational constraints different from those directly studied can lead to severe mistakes.

Finally, we explored the application of this age-estimation technique to stars in GCs and compared the results with the findings of the classical isochrone-fitting age-estimation method. As typical observational uncertainties, we adopted the ones obtained by \citet{Miglio2016} for stars in M4, the only GC for which asteroseismic data are available.
The analysis with synthetic stars show that, as the uncertainties in effective temperature and asteroseismic parameters are relatively large, the expected precision of age estimates is around 20\% for MS stars and up to 40\% for RGB stars. Classical GC age-estimation methods such as the vertical method, can achieve significantly better accuracy. A more accurate comparison of the precision achievable by classical and asteroseismology-based methods would involve adopting MCMC methods and explicitly imposing a common age in the fit of asteroseismic data. However, this comparison is outside the scope of the present paper.

To empirically explore the impact of systematic discrepancies between models and real data, we applied the asteroseismic age-estimation method to seven RGB stars in M4, for which high-quality asteroseismic observation are available \citep{Miglio2016}, repeating the procedure already carried out by \citep{Miglio2016} and \citet{Malavolta14}.
To this end, we recomputed the seven stellar effective temperatures using photometric data from \citet{Stetson2019} and the third early data release of Gaia \citep{EDR3phot}. This procedure was necessary because the temperatures from \citet{Marino2008}, adopted by \citep{Miglio2016}, did not satisfactory agree with those of the isochrone that best fitted the colour--magnitude diagram of the cluster. Our estimates of effective temperatures are generally 100 K higher than those in \citet{Marino2008}, but they are in good agreement with the values from \citet{Malavolta14} and are in better agreement with extrapolated temperatures from the theoretical isochrone that fitted the CMD.

We also applied a correction to the scaling relations for the asteroseismic parameters as the analysed M4 stars are all in the RGB phase, close to the RGB bump, where these corrections are important.
The asteroseismic age and mass at the TO ($11.9 \pm 1.5$ Gyr and $M=0.86 \pm 0.04$ $M_{\sun}$, respectively) are in good agreement with literature results. However, uncertainties in estimated parameters are relatively large, as expected from the results of the theoretical analysis.

\begin{acknowledgements}
We thank our anonymous referee for the useful comments and suggestions. 
P.G.P.M. and S.D. acknowledge INFN (Iniziativa specifica TAsP). 
\end{acknowledgements}

\bibliographystyle{aa}
\bibliography{biblio}

\begin{thebibliography}{87}
\expandafter\ifx\csname natexlab\endcsname\relax\def\natexlab#1{#1}\fi

\bibitem[{{Appourchaux} {et~al.}(2008){Appourchaux}, {Michel}, {Auvergne},
  {Baglin}, {Toutain}, {Baudin}, {Benomar}, {Chaplin}, {Deheuvels}, {Samadi},
  {Verner}, {Boumier}, {Garc{\'{\i}}a}, {Mosser}, {Hulot}, {Ballot}, {Barban},
  {Elsworth}, {Jim{\'e}nez-Reyes}, {Kjeldsen}, {R{\'e}gulo}, \&
  {Roxburgh}}]{Appourchaux2008}
{Appourchaux}, T., {Michel}, E., {Auvergne}, M., {et~al.} 2008, \aap, 488, 705

\bibitem[{{Asplund} {et~al.}(2009){Asplund}, {Grevesse}, {Sauval}, \&
  {Scott}}]{AGSS09}
{Asplund}, M., {Grevesse}, N., {Sauval}, A.~J., \& {Scott}, P. 2009, \araa, 47,
  481

\bibitem[{{Baglin} {et~al.}(2009){Baglin}, {Auvergne}, {Barge}, {Deleuil},
  {Michel}, \& {CoRoT Exoplanet Science Team}}]{Baglin2009}
{Baglin}, A., {Auvergne}, M., {Barge}, P., {et~al.} 2009, in IAU Symposium,
  Vol. 253, IAU Symposium, ed. F.~{Pont}, D.~{Sasselov}, \& M.~J. {Holman},
  71--81

\bibitem[{{Bahcall} {et~al.}(2001){Bahcall}, {Pinsonneault}, \&
  {Basu}}]{Bahcall2001}
{Bahcall}, J.~N., {Pinsonneault}, M.~H., \& {Basu}, S. 2001, \apj, 555, 990

\bibitem[{{Bailin}(2019)}]{Bailin2019}
{Bailin}, J. 2019, \apjs, 245, 5

\bibitem[{{Basu} {et~al.}(2010){Basu}, {Chaplin}, \& {Elsworth}}]{Basu2010}
{Basu}, S., {Chaplin}, W.~J., \& {Elsworth}, Y. 2010, \apj, 710, 1596

\bibitem[{{Basu} {et~al.}(2012){Basu}, {Verner}, {Chaplin}, \&
  {Elsworth}}]{Basu2012}
{Basu}, S., {Verner}, G.~A., {Chaplin}, W.~J., \& {Elsworth}, Y. 2012, \apj,
  746, 76

\bibitem[{{Bonaca} {et~al.}(2012){Bonaca}, {Tanner}, {Basu}, {Chaplin},
  {Metcalfe}, {Monteiro}, {Ballot}, {Bedding}, {Bonanno}, {Broomhall},
  {Bruntt}, {Campante}, {Christensen-Dalsgaard}, {Corsaro}, {Elsworth},
  {Garc{\'{\i}}a}, {Hekker}, {Karoff}, {Kjeldsen}, {Mathur}, {R{\'e}gulo},
  {Roxburgh}, {Stello}, {Trampedach}, {Barclay}, {Burke}, \&
  {Caldwell}}]{Bonaca2012}
{Bonaca}, A., {Tanner}, J.~D., {Basu}, S., {et~al.} 2012, \apjl, 755, L12

\bibitem[{{Borucki} {et~al.}(2010){Borucki}, {Koch}, {Basri}, {Batalha},
  {Brown}, {Caldwell}, {Caldwell}, {Christensen-Dalsgaard}, {Cochran},
  {DeVore}, {Dunham}, {Dupree}, {Gautier}, {Geary}, {Gilliland}, {Gould},
  {Howell}, {Jenkins}, {Kondo}, {Latham}, {Marcy}, {Meibom}, {Kjeldsen},
  {Lissauer}, {Monet}, {Morrison}, {Sasselov}, {Tarter}, {Boss}, {Brownlee},
  {Owen}, {Buzasi}, {Charbonneau}, {Doyle}, {Fortney}, {Ford}, {Holman},
  {Seager}, {Steffen}, {Welsh}, {Rowe}, {Anderson}, {Buchhave}, {Ciardi},
  {Walkowicz}, {Sherry}, {Horch}, {Isaacson}, {Everett}, {Fischer}, {Torres},
  {Johnson}, {Endl}, {MacQueen}, {Bryson}, {Dotson}, {Haas}, {Kolodziejczak},
  {Van Cleve}, {Chandrasekaran}, {Twicken}, {Quintana}, {Clarke}, {Allen},
  {Li}, {Wu}, {Tenenbaum}, {Verner}, {Bruhweiler}, {Barnes}, \&
  {Prsa}}]{Borucki2010}
{Borucki}, W.~J., {Koch}, D., {Basri}, G., {et~al.} 2010, Science, 327, 977

\bibitem[{{Brown} {et~al.}(1991){Brown}, {Gilliland}, {Noyes}, \&
  {Ramsey}}]{Brown1991}
{Brown}, T.~M., {Gilliland}, R.~L., {Noyes}, R.~W., \& {Ramsey}, L.~W. 1991,
  \apj, 368, 599

\bibitem[{{Caputo} {et~al.}(1985)}]{Caputo_85}
{Caputo}, F. {et~al.} 1985, aap, 143, 8

\bibitem[{{Casagrande} {et~al.}(2014){Casagrande}, {Silva Aguirre}, {Stello},
  {Huber}, {Serenelli}, {Cassisi}, {Dotter}, {Milone}, {Hodgkin}, {Marino},
  {Lund}, {Pietrinferni}, {Asplund}, {Feltzing}, {Flynn}, {Grundahl}, {Nissen},
  {Sch{\"o}nrich}, {Schlesinger}, \& {Wang}}]{Casagrande2014}
{Casagrande}, L., {Silva Aguirre}, V., {Stello}, D., {et~al.} 2014, \apj, 787,
  110

\bibitem[{{Chaboyer} {et~al.}(2001){Chaboyer}, {Fenton}, {Nelan}, {Patnaude},
  \& {Simon}}]{Chaboyer2001}
{Chaboyer}, B., {Fenton}, W.~H., {Nelan}, J.~E., {Patnaude}, D.~J., \& {Simon},
  F.~E. 2001, \apj, 562, 521

\bibitem[{{Chaplin} {et~al.}(2014){Chaplin}, {Basu}, {Huber}, {Serenelli},
  {Casagrande}, {Silva Aguirre}, {Ball}, {Creevey}, {Gizon}, {Handberg},
  {Karoff}, {Lutz}, {Marques}, {Miglio}, {Stello}, {Suran}, {Pricopi},
  {Metcalfe}, {Monteiro}, {Molenda-{\.Z}akowicz}, {Appourchaux},
  {Christensen-Dalsgaard}, {Elsworth}, {Garc{\'{\i}}a}, {Houdek}, {Kjeldsen},
  {Bonanno}, {Campante}, {Corsaro}, {Gaulme}, {Hekker}, {Mathur}, {Mosser},
  {R{\'e}gulo}, \& {Salabert}}]{Chaplin2014}
{Chaplin}, W.~J., {Basu}, S., {Huber}, D., {et~al.} 2014, \apjs, 210, 1

\bibitem[{{Clausen} {et~al.}(2009){Clausen}, {Bruntt}, {Claret}, {Larsen},
  {Andersen}, {Nordstr{\"o}m}, \& {Gim{\'e}nez}}]{Clausen2009}
{Clausen}, J.~V., {Bruntt}, H., {Claret}, A., {et~al.} 2009, \aap, 502, 253

\bibitem[{{Das} {et~al.}(2020){Das}, {Hawkins}, \& {Jofr{\'e}}}]{Das2020}
{Das}, P., {Hawkins}, K., \& {Jofr{\'e}}, P. 2020, \mnras, 493, 5195

\bibitem[{{Degl'Innocenti} {et~al.}(2008){Degl'Innocenti}, {Prada Moroni},
  {Marconi}, \& {Ruoppo}}]{scilla2008}
{Degl'Innocenti}, S., {Prada Moroni}, P.~G., {Marconi}, M., \& {Ruoppo}, A.
  2008, \apss, 316, 25

\bibitem[{{Deheuvels} \& {Michel}(2011)}]{Deheuvels2011}
{Deheuvels}, S. \& {Michel}, E. 2011, \aap, 535, A91

\bibitem[{Dell'Omodarme \& Valle(2013)}]{stellar}
Dell'Omodarme, M. \& Valle, G. 2013, The R Journal, 5, 108

\bibitem[{{Dell'Omodarme} {et~al.}(2012){Dell'Omodarme}, {Valle},
  {Degl'Innocenti}, \& {Prada Moroni}}]{database2012}
{Dell'Omodarme}, M., {Valle}, G., {Degl'Innocenti}, S., \& {Prada Moroni},
  P.~G. 2012, A\&A, 540, A26

\bibitem[{{Dotter} {et~al.}(2010{\natexlab{a}}){Dotter}, {Sarajedini},
  {Anderson}, {Aparicio}, {Bedin}, {Chaboyer}, {Majewski}, {Mar{\'\i}n-Franch},
  {Milone}, {Paust}, {Piotto}, {Reid}, {Rosenberg}, \& {Siegel}}]{Dotter2010}
{Dotter}, A., {Sarajedini}, A., {Anderson}, J., {et~al.} 2010{\natexlab{a}},
  \apj, 708, 698

\bibitem[{{Dotter} {et~al.}(2010{\natexlab{b}})}]{Dotter}
{Dotter}, A. {et~al.} 2010{\natexlab{b}}, apj, 708, 698

\bibitem[{{Epstein} {et~al.}(2014){Epstein}, {Elsworth}, {Johnson}, {Shetrone},
  {Mosser}, {Hekker}, {Tayar}, {Harding}, {Pinsonneault}, {Silva Aguirre},
  {Basu}, {Beers}, {Bizyaev}, {Bedding}, {Chaplin}, {Frinchaboy}, {Garc{'\i}a},
  {Garc{'\i}a P{'e}rez}, {Hearty}, {Huber}, {Ivans}, {Majewski}, {Mathur},
  {Nidever}, {Serenelli}, {Schiavon}, {Schneider}, {Sch{"o}nrich}, {Sobeck},
  {Stassun}, {Stello}, \& {Zasowski}}]{Epstein2014b}
{Epstein}, C.~R., {Elsworth}, Y.~P., {Johnson}, J.~A., {et~al.} 2014, \apjl,
  785, L28

\bibitem[{{Gai} {et~al.}(2011){Gai}, {Basu}, {Chaplin}, \&
  {Elsworth}}]{Gai2011}
{Gai}, N., {Basu}, S., {Chaplin}, W.~J., \& {Elsworth}, Y. 2011, \apj, 730, 63

\bibitem[{{Gaulme} {et~al.}(2016){Gaulme}, {McKeever}, {Jackiewicz}, {Rawls},
  {Corsaro}, {Mosser}, {Southworth}, {Mahadevan}, {Bender}, \&
  {Deshpande}}]{Gaulme2016}
{Gaulme}, P., {McKeever}, J., {Jackiewicz}, J., {et~al.} 2016, \apj, 832, 121

\bibitem[{{Gilliland} {et~al.}(2010){Gilliland}, {Brown},
  {Christensen-Dalsgaard}, {Kjeldsen}, {Aerts}, {Appourchaux}, {Basu},
  {Bedding}, {Chaplin}, {Cunha}, {De Cat}, {De Ridder}, {Guzik}, {Handler},
  {Kawaler}, {Kiss}, {Kolenberg}, {Kurtz}, {Metcalfe}, {Monteiro}, {Szab{\'o}},
  {Arentoft}, {Balona}, {Debosscher}, {Elsworth}, {Quirion}, {Stello},
  {Su{\'a}rez}, {Borucki}, {Jenkins}, {Koch}, {Kondo}, {Latham}, {Rowe}, \&
  {Steffen}}]{Gilliland2010}
{Gilliland}, R.~L., {Brown}, T.~M., {Christensen-Dalsgaard}, J., {et~al.} 2010,
  \pasp, 122, 131

\bibitem[{{Gratton} {et~al.}(2011){Gratton}, {Johnson}, {Lucatello}, {D'Orazi},
  \& {Pilachowski}}]{Gratton2011}
{Gratton}, R.~G., {Johnson}, C.~I., {Lucatello}, S., {D'Orazi}, V., \&
  {Pilachowski}, C. 2011, \aap, 534, A72

\bibitem[{{Grunblatt} {et~al.}(2021){Grunblatt}, {Zinn}, {Price-Whelan},
  {Angus}, {Saunders}, {Hon}, {Stokholm}, {Bellinger}, {Martell}, {Mosser},
  {Cunningham}, {Tayar}, {Huber}, {R{\o}rsted}, \& {Silva
  Aguirre}}]{Grunblatt2021}
{Grunblatt}, S.~K., {Zinn}, J.~C., {Price-Whelan}, A.~M., {et~al.} 2021, \apj,
  916, 88

\bibitem[{{Gruyters} {et~al.}(2014){Gruyters}, {Nordlander}, \&
  {Korn}}]{Gruyters2014}
{Gruyters}, P., {Nordlander}, T., \& {Korn}, A.~J. 2014, \aap, 567, A72

\bibitem[{{Guo} {et~al.}(2016){Guo}, {Liu}, \& {Liu}}]{Guo2016}
{Guo}, J.-C., {Liu}, C., \& {Liu}, J.-F. 2016, Research in Astronomy and
  Astrophysics, 16, 44

\bibitem[{{Hansen} {et~al.}(2002)}]{Hansen_2002}
{Hansen}, B. M.~S. {et~al.} 2002, apjl, 574, L155

\bibitem[{{Hendricks} {et~al.}(2012){Hendricks}, {Stetson}, {VandenBerg}, \&
  {Dall'Ora}}]{Hendricks2012}
{Hendricks}, B., {Stetson}, P.~B., {VandenBerg}, D.~A., \& {Dall'Ora}, M. 2012,
  \aj, 144, 25

\bibitem[{{Jofr{\'e}} \& {Weiss}(2011)}]{Jofre2011}
{Jofr{\'e}}, P. \& {Weiss}, A. 2011, \aap, 533, A59

\bibitem[{{J{\o}rgensen} \& {Lindegren}(2005)}]{jorgensen2005}
{J{\o}rgensen}, B.~R. \& {Lindegren}, L. 2005, \aap, 436, 127

\bibitem[{{Kjeldsen} \& {Bedding}(1995)}]{Kjeldsen1995}
{Kjeldsen}, H. \& {Bedding}, T.~R. 1995, \aap, 293, 87

\bibitem[{{Korn} {et~al.}(2007){Korn}, {Grundahl}, {Richard}, {Mashonkina},
  {Barklem}, {Collet}, {Gustafsson}, \& {Piskunov}}]{Korn2007}
{Korn}, A.~J., {Grundahl}, F., {Richard}, O., {et~al.} 2007, \apj, 671, 402

\bibitem[{{Kroupa}(2002)}]{Kroupa2002}
{Kroupa}, P. 2002, Science, 295, 82

\bibitem[{{Lebreton} {et~al.}(2014){Lebreton}, {Goupil}, \&
  {Montalb{\'a}n}}]{Lebreton2014}
{Lebreton}, Y., {Goupil}, M.~J., \& {Montalb{\'a}n}, J. 2014, in EAS
  Publications Series, Vol.~65, EAS Publications Series, 99--176

\bibitem[{{Magic} {et~al.}(2015){Magic}, {Weiss}, \& {Asplund}}]{Magic2014}
{Magic}, Z., {Weiss}, A., \& {Asplund}, M. 2015, \aap, 573, A89

\bibitem[{{Malavolta} {et~al.}(2014){Malavolta}, {Sneden}, {Piotto}, {Milone},
  {Bedin}, \& {Nascimbeni}}]{Malavolta14}
{Malavolta}, L., {Sneden}, C., {Piotto}, G., {et~al.} 2014, \aj, 147, 25

\bibitem[{{Marino} {et~al.}(2008){Marino}, {Villanova}, {Piotto}, {Milone},
  {Momany}, {Bedin}, \& {Medling}}]{Marino2008}
{Marino}, A.~F., {Villanova}, S., {Piotto}, G., {et~al.} 2008, \aap, 490, 625

\bibitem[{{Mathur} {et~al.}(2012){Mathur}, {Metcalfe}, {Woitaszek}, {Bruntt},
  {Verner}, {Christensen-Dalsgaard}, {Creevey}, {Do{\v g}an}, {Basu}, {Karoff},
  {Stello}, {Appourchaux}, {Campante}, {Chaplin}, {Garc{\'{\i}}a}, {Bedding},
  {Benomar}, {Bonanno}, {Deheuvels}, {Elsworth}, {Gaulme}, {Guzik}, {Handberg},
  {Hekker}, {Herzberg}, {Monteiro}, {Piau}, {Quirion}, {R{\'e}gulo}, {Roth},
  {Salabert}, {Serenelli}, {Thompson}, {Trampedach}, {White}, {Ballot},
  {Brand{\~a}o}, {Molenda-{\.Z}akowicz}, {Kjeldsen}, {Twicken}, {Uddin}, \&
  {Wohler}}]{Mathur2012}
{Mathur}, S., {Metcalfe}, T.~S., {Woitaszek}, M., {et~al.} 2012, \apj, 749, 152

\bibitem[{{Matsuno} {et~al.}(2021){Matsuno}, {Aoki}, {Casagrande}, {Ishigaki},
  {Shi}, {Takata}, {Xiang}, {Yong}, {Li}, {Suda}, {Xing}, \&
  {Zhao}}]{Matsuno2021}
{Matsuno}, T., {Aoki}, W., {Casagrande}, L., {et~al.} 2021, \apj, 912, 72

\bibitem[{{Michel} {et~al.}(2008){Michel}, {Baglin}, {Auvergne}, {Catala},
  {Samadi}, {Baudin}, {Appourchaux}, {Barban}, {Weiss}, {Berthomieu},
  {Boumier}, {Dupret}, {Garcia}, {Fridlund}, {Garrido}, {Goupil}, {Kjeldsen},
  {Lebreton}, {Mosser}, {Grotsch-Noels}, {Janot-Pacheco}, {Provost},
  {Roxburgh}, {Thoul}, {Toutain}, {Tiph{\`e}ne}, {Turck-Chieze}, {Vauclair},
  {Vauclair}, {Aerts}, {Alecian}, {Ballot}, {Charpinet}, {Hubert},
  {Ligni{\`e}res}, {Mathias}, {Monteiro}, {Neiner}, {Poretti}, {Renan de
  Medeiros}, {Ribas}, {Rieutord}, {Cort{\'e}s}, \& {Zwintz}}]{Michel2008}
{Michel}, E., {Baglin}, A., {Auvergne}, M., {et~al.} 2008, Science, 322, 558

\bibitem[{{Miglio} {et~al.}(2016){Miglio}, {Chaplin}, {Brogaard}, {Lund},
  {Mosser}, {Davies}, {Handberg}, {Milone}, {Marino}, {Bossini}, {Elsworth},
  {Grundahl}, {Arentoft}, {Bedin}, {Campante}, {Jessen-Hansen}, {Jones},
  {Kuszlewicz}, {Malavolta}, {Nascimbeni}, \& {Sandquist}}]{Miglio2016}
{Miglio}, A., {Chaplin}, W.~J., {Brogaard}, K., {et~al.} 2016, \mnras, 461, 760

\bibitem[{{Miglio} {et~al.}(2021){Miglio}, {Girardi}, {Grundahl}, {Mosser},
  {Bastian}, {Bragaglia}, {Brogaard}, {Buldgen}, {Chantereau}, {Chaplin},
  {Chiappini}, {Dupret}, {Eggenberger}, {Gieles}, {Izzard}, {Kawata}, {Karoff},
  {Lagarde}, {Mackereth}, {Magrin}, {Meynet}, {Michel}, {Montalb{\'a}n},
  {Nascimbeni}, {Noels}, {Piotto}, {Ragazzoni}, {Soszy{\'n}ski}, {Tolstoy},
  {Toonen}, {Triaud}, \& {Vincenzo}}]{Miglio2021}
{Miglio}, A., {Girardi}, L., {Grundahl}, F., {et~al.} 2021, Experimental
  Astronomy, 51, 963

\bibitem[{{Montalb{\'a}n} {et~al.}(2021){Montalb{\'a}n}, {Mackereth}, {Miglio},
  {Vincenzo}, {Chiappini}, {Buldgen}, {Mosser}, {Noels}, {Scuflaire}, {Vrard},
  {Willett}, {Davies}, {Hall}, {Nielsen}, {Khan}, {Rendle}, {van Rossem},
  {Ferguson}, \& {Chaplin}}]{Montalban2021}
{Montalb{\'a}n}, J., {Mackereth}, J.~T., {Miglio}, A., {et~al.} 2021, Nature
  Astronomy, 5, 640

\bibitem[{{Mucciarelli} \& {Bellazzini}(2020)}]{Mucciarelli2020}
{Mucciarelli}, A. \& {Bellazzini}, M. 2020, Research Notes of the American
  Astronomical Society, 4, 52

\bibitem[{{Nordlander} {et~al.}(2012){Nordlander}, {Korn}, {Richard}, \&
  {Lind}}]{Nordlander2012}
{Nordlander}, T., {Korn}, A.~J., {Richard}, O., \& {Lind}, K. 2012, \apj, 753,
  48

\bibitem[{{Planck Collaboration} {et~al.}(2020){Planck Collaboration},
  {Aghanim}, {Akrami}, {Ashdown}, {Aumont}, {Baccigalupi}, {Ballardini},
  {Banday}, {Barreiro}, {Bartolo}, {Basak}, {Battye}, {Benabed}, {Bernard},
  {Bersanelli}, {Bielewicz}, {Bock}, {Bond}, {Borrill}, {Bouchet}, {Boulanger},
  {Bucher}, {Burigana}, {Butler}, {Calabrese}, {Cardoso}, {Carron},
  {Challinor}, {Chiang}, {Chluba}, {Colombo}, {Combet}, {Contreras}, {Crill},
  {Cuttaia}, {de Bernardis}, {de Zotti}, {Delabrouille}, {Delouis}, {Di
  Valentino}, {Diego}, {Dor{\'e}}, {Douspis}, {Ducout}, {Dupac}, {Dusini},
  {Efstathiou}, {Elsner}, {En{\ss}lin}, {Eriksen}, {Fantaye}, {Farhang},
  {Fergusson}, {Fernandez-Cobos}, {Finelli}, {Forastieri}, {Frailis},
  {Fraisse}, {Franceschi}, {Frolov}, {Galeotta}, {Galli}, {Ganga},
  {G{\'e}nova-Santos}, {Gerbino}, {Ghosh}, {Gonz{\'a}lez-Nuevo}, {G{\'o}rski},
  {Gratton}, {Gruppuso}, {Gudmundsson}, {Hamann}, {Handley}, {Hansen},
  {Herranz}, {Hildebrandt}, {Hivon}, {Huang}, {Jaffe}, {Jones}, {Karakci},
  {Keih{\"a}nen}, {Keskitalo}, {Kiiveri}, {Kim}, {Kisner}, {Knox},
  {Krachmalnicoff}, {Kunz}, {Kurki-Suonio}, {Lagache}, {Lamarre}, {Lasenby},
  {Lattanzi}, {Lawrence}, {Le Jeune}, {Lemos}, {Lesgourgues}, {Levrier},
  {Lewis}, {Liguori}, {Lilje}, {Lilley}, {Lindholm}, {L{\'o}pez-Caniego},
  {Lubin}, {Ma}, {Mac{\'\i}as-P{\'e}rez}, {Maggio}, {Maino}, {Mandolesi},
  {Mangilli}, {Marcos-Caballero}, {Maris}, {Martin}, {Martinelli},
  {Mart{\'\i}nez-Gonz{\'a}lez}, {Matarrese}, {Mauri}, {McEwen}, {Meinhold},
  {Melchiorri}, {Mennella}, {Migliaccio}, {Millea}, {Mitra},
  {Miville-Desch{\^e}nes}, {Molinari}, {Montier}, {Morgante}, {Moss}, {Natoli},
  {N{\o}rgaard-Nielsen}, {Pagano}, {Paoletti}, {Partridge}, {Patanchon},
  {Peiris}, {Perrotta}, {Pettorino}, {Piacentini}, {Polastri}, {Polenta},
  {Puget}, {Rachen}, {Reinecke}, {Remazeilles}, {Renzi}, {Rocha}, {Rosset},
  {Roudier}, {Rubi{\~n}o-Mart{\'\i}n}, {Ruiz-Granados}, {Salvati}, {Sandri},
  {Savelainen}, {Scott}, {Shellard}, {Sirignano}, {Sirri}, {Spencer},
  {Sunyaev}, {Suur-Uski}, {Tauber}, {Tavagnacco}, {Tenti}, {Toffolatti},
  {Tomasi}, {Trombetti}, {Valenziano}, {Valiviita}, {Van Tent}, {Vibert},
  {Vielva}, {Villa}, {Vittorio}, {Wandelt}, {Wehus}, {White}, {White},
  {Zacchei}, \& {Zonca}}]{Planck2020}
{Planck Collaboration}, {Aghanim}, N., {Akrami}, Y., {et~al.} 2020, \aap, 641,
  A6

\bibitem[{{Ram{\'{\i}}rez} \& {Mel{\'e}ndez}(2005)}]{Ramirez2005}
{Ram{\'{\i}}rez}, I. \& {Mel{\'e}ndez}, J. 2005, \apj, 626, 465

\bibitem[{{Ricker} {et~al.}(2015){Ricker}, {Winn}, {Vanderspek}, {Latham},
  {Bakos}, {Bean}, {Berta-Thompson}, {Brown}, {Buchhave}, {Butler}, {Butler},
  {Chaplin}, {Charbonneau}, {Christensen-Dalsgaard}, {Clampin}, {Deming},
  {Doty}, {De Lee}, {Dressing}, {Dunham}, {Endl}, {Fressin}, {Ge}, {Henning},
  {Holman}, {Howard}, {Ida}, {Jenkins}, {Jernigan}, {Johnson}, {Kaltenegger},
  {Kawai}, {Kjeldsen}, {Laughlin}, {Levine}, {Lin}, {Lissauer}, {MacQueen},
  {Marcy}, {McCullough}, {Morton}, {Narita}, {Paegert}, {Palle}, {Pepe},
  {Pepper}, {Quirrenbach}, {Rinehart}, {Sasselov}, {Sato}, {Seager},
  {Sozzetti}, {Stassun}, {Sullivan}, {Szentgyorgyi}, {Torres}, {Udry}, \&
  {Villasenor}}]{Ricker2015}
{Ricker}, G.~R., {Winn}, J.~N., {Vanderspek}, R., {et~al.} 2015, Journal of
  Astronomical Telescopes, Instruments, and Systems, 1, 014003

\bibitem[{{Riello} {et~al.}(2020){Riello}, {De Angeli}, {Evans}, {Montegriffo},
  {Carrasco}, {Busso}, {Palaversa}, {Burgess}, {Diener}, {Davidson}, {Rowell},
  {Fabricius}, {Jordi}, {Bellazzini}, {Pancino}, {Harrison}, {Cacciari}, {van
  Leeuwen}, {Hambly}, {Hodgkin}, {Osborne}, {Altavilla}, {Barstow}, {Brown},
  {Castellani}, {Cowell}, {De Luise}, {Gilmore}, {Giuffrida}, {Hidalgo},
  {Holland}, {Marinoni}, {Pagani}, {Piersimoni}, {Pulone}, {Ragaini}, {Rainer},
  {Richards}, {Sanna}, {Walton}, {Weiler}, \& {Yoldas}}]{EDR3phot}
{Riello}, M., {De Angeli}, F., {Evans}, D.~W., {et~al.} 2020, arXiv e-prints,
  arXiv:2012.01916

\bibitem[{{Rodrigues} {et~al.}(2017){Rodrigues}, {Bossini}, {Miglio},
  {Girardi}, {Montalb{\'a}n}, {Noels}, {Trabucchi}, {Coelho}, \&
  {Marigo}}]{Rodrigued2017}
{Rodrigues}, T.~S., {Bossini}, D., {Miglio}, A., {et~al.} 2017, \mnras, 467,
  1433

\bibitem[{{Salaris} \& {Weiss}(2002{\natexlab{a}})}]{Salaris_02}
{Salaris}, M. \& {Weiss}, A. 2002{\natexlab{a}}, aap, 388, 492

\bibitem[{{Salaris} \& {Weiss}(2002{\natexlab{b}})}]{salaris2002}
{Salaris}, M. \& {Weiss}, A. 2002{\natexlab{b}}, \aap, 388, 492

\bibitem[{{Salpeter}(1955)}]{Salpeter1955}
{Salpeter}, E.~E. 1955, \apj, 121, 161

\bibitem[{{Sanders} \& {Das}(2018)}]{Sanders2018}
{Sanders}, J.~L. \& {Das}, P. 2018, \mnras, 481, 4093

\bibitem[{{Sandquist} {et~al.}(2016){Sandquist}, {Jessen-Hansen}, {Shetrone},
  {Brogaard}, {Meibom}, {Leitner}, {Stello}, {Bruntt}, {Antoci}, {Orosz},
  {Grundahl}, \& {Frandsen}}]{Sandquist2016}
{Sandquist}, E.~L., {Jessen-Hansen}, J., {Shetrone}, M.~D., {et~al.} 2016,
  \apj, 831, 11

\bibitem[{{Sharma} {et~al.}(2016){Sharma}, {Stello}, {Bland-Hawthorn}, {Huber},
  \& {Bedding}}]{Sharma2016}
{Sharma}, S., {Stello}, D., {Bland-Hawthorn}, J., {Huber}, D., \& {Bedding},
  T.~R. 2016, \apj, 822, 15

\bibitem[{{Silva Aguirre} {et~al.}(2017){Silva Aguirre}, {Lund}, {Antia},
  {Ball}, {Basu}, {Christensen-Dalsgaard}, {Lebreton}, {Reese}, {Verma},
  {Casagrande}, {Justesen}, {Mosumgaard}, {Chaplin}, {Bedding}, {Davies},
  {Handberg}, {Houdek}, {Huber}, {Kjeldsen}, {Latham}, {White}, {Coelho},
  {Miglio}, \& {Rendle}}]{SilvaAguirre2017}
{Silva Aguirre}, V., {Lund}, M.~N., {Antia}, H.~M., {et~al.} 2017, \apj, 835,
  173

\bibitem[{{Soderblom}(2010)}]{Soderblom2010}
{Soderblom}, D.~R. 2010, \araa, 48, 581

\bibitem[{{Stello} {et~al.}(2022){Stello}, {Saunders}, {Grunblatt}, {Hon},
  {Reyes}, {Huber}, {Bedding}, {Elsworth}, {Garc{'\i}a}, {Hekker}, {Kallinger},
  {Mathur}, {Mosser}, \& {Pinsonneault}}]{Stello2022}
{Stello}, D., {Saunders}, N., {Grunblatt}, S., {et~al.} 2022, \mnras, 512, 1677

\bibitem[{{Stetson} {et~al.}(2019){Stetson}, {Pancino}, {Zocchi}, {Sanna}, \&
  {Monelli}}]{Stetson2019}
{Stetson}, P.~B., {Pancino}, E., {Zocchi}, A., {Sanna}, N., \& {Monelli}, M.
  2019, \mnras, 485, 3042

\bibitem[{{Tailo} {et~al.}(2022){Tailo}, {Corsaro}, {Miglio}, {Montalb{\'a}n},
  {Brogaard}, {Milone}, {Stokholm}, {Casali}, \& {Bragaglia}}]{Tail02022}
{Tailo}, M., {Corsaro}, E., {Miglio}, A., {et~al.} 2022, \aap, 662, L7

\bibitem[{{Takeda} {et~al.}(2007){Takeda}, {Ford}, {Sills}, {Rasio}, {Fischer},
  \& {Valenti}}]{Takeda07}
{Takeda}, G., {Ford}, E.~B., {Sills}, A., {et~al.} 2007, \apjs, 168, 297

\bibitem[{{Tanner} {et~al.}(2014){Tanner}, {Basu}, \& {Demarque}}]{Tanner2014}
{Tanner}, J.~D., {Basu}, S., \& {Demarque}, P. 2014, \apjl, 785, L13

\bibitem[{{Tayar} {et~al.}(2022){Tayar}, {Claytor}, {Huber}, \& {van
  Saders}}]{Tayar2022}
{Tayar}, J., {Claytor}, Z.~R., {Huber}, D., \& {van Saders}, J. 2022, \apj,
  927, 31

\bibitem[{{Thoul} {et~al.}(1994){Thoul}, {Bahcall}, \& {Loeb}}]{thoul94}
{Thoul}, A.~A., {Bahcall}, J.~N., \& {Loeb}, A. 1994, \apj, 421, 828

\bibitem[{{Tognelli} {et~al.}(2021){Tognelli}, {Dell'Omodarme}, {Valle}, {Prada
  Moroni}, \& {Degl'Innocenti}}]{Tognelli2021}
{Tognelli}, E., {Dell'Omodarme}, M., {Valle}, G., {Prada Moroni}, P.~G., \&
  {Degl'Innocenti}, S. 2021, \mnras, 501, 383

\bibitem[{{Trampedach} \& {Stein}(2011)}]{Trampedach2011}
{Trampedach}, R. \& {Stein}, R.~F. 2011, \apj, 731, 78

\bibitem[{{Ulrich}(1986)}]{Ulrich1986}
{Ulrich}, R.~K. 1986, \apjl, 306, L37

\bibitem[{{Valle} {et~al.}(2014){Valle}, {Dell'Omodarme}, {Prada Moroni}, \&
  {Degl'Innocenti}}]{scepter1}
{Valle}, G., {Dell'Omodarme}, M., {Prada Moroni}, P.~G., \& {Degl'Innocenti},
  S. 2014, \aap, 561, A125

\bibitem[{{Valle} {et~al.}(2015{\natexlab{a}}){Valle}, {Dell'Omodarme}, {Prada
  Moroni}, \& {Degl'Innocenti}}]{binary}
{Valle}, G., {Dell'Omodarme}, M., {Prada Moroni}, P.~G., \& {Degl'Innocenti},
  S. 2015{\natexlab{a}}, \aap, 579, A59

\bibitem[{{Valle} {et~al.}(2015{\natexlab{b}}){Valle}, {Dell'Omodarme}, {Prada
  Moroni}, \& {Degl'Innocenti}}]{bulge}
{Valle}, G., {Dell'Omodarme}, M., {Prada Moroni}, P.~G., \& {Degl'Innocenti},
  S. 2015{\natexlab{b}}, \aap, 577, A72

\bibitem[{{Valle} {et~al.}(2015{\natexlab{c}}){Valle}, {Dell'Omodarme}, {Prada
  Moroni}, \& {Degl'Innocenti}}]{eta}
{Valle}, G., {Dell'Omodarme}, M., {Prada Moroni}, P.~G., \& {Degl'Innocenti},
  S. 2015{\natexlab{c}}, \aap, 575, A12

\bibitem[{{Valle} {et~al.}(2018{\natexlab{a}}){Valle}, {Dell'Omodarme}, {Prada
  Moroni}, \& {Degl'Innocenti}}]{Valle2018}
{Valle}, G., {Dell'Omodarme}, M., {Prada Moroni}, P.~G., \& {Degl'Innocenti},
  S. 2018{\natexlab{a}}, \aap, 620, A168

\bibitem[{{Valle} {et~al.}(2020){Valle}, {Dell'Omodarme}, {Prada Moroni}, \&
  {Degl'Innocenti}}]{smallsep}
{Valle}, G., {Dell'Omodarme}, M., {Prada Moroni}, P.~G., \& {Degl'Innocenti},
  S. 2020, \aap, 635, A77

\bibitem[{{Valle} {et~al.}(2018{\natexlab{b}}){Valle}, {Dell'Omodarme},
  {Tognelli}, {Prada Moroni}, \& {Degl'Innocenti}}]{cluster2018}
{Valle}, G., {Dell'Omodarme}, M., {Tognelli}, E., {Prada Moroni}, P.~G., \&
  {Degl'Innocenti}, S. 2018{\natexlab{b}}, \aap, 619, A158

\bibitem[{{Valle} {et~al.}(2009){Valle}, {Marconi}, {Degl'Innocenti}, \& {Prada
  Moroni}}]{cefeidi}
{Valle}, G., {Marconi}, M., {Degl'Innocenti}, S., \& {Prada Moroni}, P.~G.
  2009, \aap, 507, 1541

\bibitem[{{VandenBerg} {et~al.}(2013{\natexlab{a}}){VandenBerg}, {Brogaard},
  {Leaman}, \& {Casagrande}}]{Vandenberg2013}
{VandenBerg}, D.~A., {Brogaard}, K., {Leaman}, R., \& {Casagrande}, L.
  2013{\natexlab{a}}, \apj, 775, 134

\bibitem[{{VandenBerg} {et~al.}(2013{\natexlab{b}})}]{Vandenberg_2013}
{VandenBerg}, D.~A. {et~al.} 2013{\natexlab{b}}, apj, 775, 134

\bibitem[{{Viani} {et~al.}(2017){Viani}, {Basu}, {Chaplin}, {Davies}, \&
  {Elsworth}}]{Viani2017}
{Viani}, L.~S., {Basu}, S., {Chaplin}, W.~J., {Davies}, G.~R., \& {Elsworth},
  Y. 2017, \apj, 843, 11

\bibitem[{{Wagner-Kaiser} {et~al.}(2017)}]{Wagner_17}
{Wagner-Kaiser}, R. {et~al.} 2017, mnras, 468, 1038

\bibitem[{{White} {et~al.}(2011){White}, {Bedding}, {Stello},
  {Christensen-Dalsgaard}, {Huber}, \& {Kjeldsen}}]{White2011}
{White}, T.~R., {Bedding}, T.~R., {Stello}, D., {et~al.} 2011, \apj, 743, 161

\bibitem[{{Y{\i}ld{\i}z}(2007)}]{Yildiz2007}
{Y{\i}ld{\i}z}, M. 2007, \mnras, 374, 1264

\bibitem[{{Yu} {et~al.}(2018){Yu}, {Huber}, {Bedding}, {Stello}, {Hon},
  {Murphy}, \& {Khanna}}]{Yu2018}
{Yu}, J., {Huber}, D., {Bedding}, T.~R., {et~al.} 2018, \apjs, 236, 42

\end{thebibliography}

\appendix

\section{Tables and likelihood maps}

\begin{table*}[]
        \centering
        \caption{SCEPtER median ($q_{50}$) and $1 \sigma$ envelope boundaries
                ($q_{16}$ and $q_{84}$) for the relative error on
age as a 
                function of the mass of the star in MS. Values
                are expressed as percent.} 
        \label{tab:global_MS}
        \begin{tabular}{lccccccccc}
                \hline\hline
                \multicolumn{10}{c}{Main Sequence}\\
                \hline\hline
                & \multicolumn{9}{c}{Mass ($M_{\sun}$)}\\
                & 0.70 & 0.72 & 0.74 & 0.76 & 0.78 & 0.80 & 0.82 & 0.84 & 0.86\\
                \hline
                \multicolumn{10}{c}{C1}\\
                $q_{16}$ & -22.5 & -23.0 & -20.8 & -18.8 & -17.2 & -16.4 & -16.8 & -17.2 & -17.3 \\ 
                $q_{50}$ & -3.8 & -1.7 & -0.3 & 0.3 & 0.0 & 0.0 & 0.1 & 0.2 & 0.5 \\ 
                $q_{84}$ & 8.6 & 14.9 & 19.2 & 21.6 & 21.4 & 20.7 & 21.2 & 22.5 & 24.2 \\ 
                \hline
                \multicolumn{10}{c}{C2}\\
                $q_{16}$ &      -22.5 & -20.8 & -19.2 & -17.6 & -16.1 & -15.4 & -16.0 & -16.8 & -17.6 \\ 
                $q_{50}$ & -2.5 & -1.2 & -0.3 & 0.1 & 0.1 & 0.1 & -0.2 & -0.6 & -1.3 \\ 
                $q_{84}$ & 6.4 & 12.9 & 17.4 & 19.9 & 19.4 & 18.6 & 19.0 & 19.7 & 20.6 \\
                \hline
                \multicolumn{10}{c}{C3}\\
                $q_{16}$ & -32.6 & -30.5 & -28.2 & -25.9 & -23.7 & -22.7 & -24.0 & -25.3 & -25.9 \\ 
                $q_{50}$ & -4.3 & -2.3 & -0.9 & -0.0 & 0.2 & 0.1 & -0.6 & -1.6 & -2.5  \\ 
                $q_{84}$ & 6.1 & 14.2 & 20.7 & 25.8 & 27.8 & 28.1 & 28.7 & 29.8 & 31.1 \\ 
                \hline
                \multicolumn{10}{c}{$\alpha_{\rm ml}$ = 1.74}\\
                $q_{16}$ & -11.9 & -11.1 & -10.3 & -9.5 & -8.6 & -7.9 & -8.1 & -8.7 & -9.5 \\  
                $q_{50}$ &  2.6 & 5.8 & 7.7 & 8.4 & 7.7 & 7.9 & 8.6 & 9.6 & 11.1 \\ 
                $q_{84}$ & 12.1 & 19.4 & 24.9 & 28.7 & 29.5 & 29.0 & 29.0 & 29.6 & 30.7 \\ 
                \hline
                \multicolumn{10}{c}{no diffusion}\\
                $q_{16}$ & -60.9 & -57.0 & -53.5 & -50.3 & -47.8 & -45.8 & -43.6 & -39.9 & -34.4 \\ 
                $q_{50}$ & -41.4 & -39.0 & -36.9 & -35.2 & -34.0 & -32.8 & -30.5 & -26.2 & -20.0 \\
                $q_{84}$ & -20.9 & -19.7 & -18.7 & -17.9 & -17.3 & -16.2 & -13.7 & -9.4 & -3.7 \\ 
                \hline
        \end{tabular}
        \tablefoot{Typical Monte Carlo relative uncertainty on $q_{16}$ and $q_{84}$ is about
                5\%, while the absolute uncertainty on $q_{50}$ is about 0.5\%.} 
\end{table*}

\begin{table*}[]
        \centering
        \caption{Same as Table~\ref{tab:global_MS} but for SGB stars.}
        \label{tab:global_SGB}
        \begin{tabular}{lcccccccccc}
                \hline\hline
                \multicolumn{11}{c}{Sub-Giant Branch}\\
                \hline\hline
                & \multicolumn{10}{c}{Mass ($M_{\sun}$)}\\
                & 0.74 & 0.76 & 0.78 & 0.80 & 0.82 & 0.84 & 0.86 & 0.88 & 0.90 & 0.92 \\
                \hline
                \multicolumn{11}{c}{C1}\\
                $q_{16}$ & -13.2 & -13.0 & -12.9 & -12.7 & -13.0 & -13.7 & -14.4 & -14.2 & -13.3 & -11.8  \\ 
                $q_{50}$ & 0 & -0.1 & -0.1 & -0.1 & 0 & 0.1 & 0.4 & 0.9 & 1.4 & 1.8 \\ 
                $q_{84}$ & 15.8 & 15.4 & 15.3 & 15.3 & 16.0 & 17.7 & 19.8 & 20.8 & 21.1 & 20.6  \\ 
                \hline
                \multicolumn{11}{c}{C2}\\
                $q_{16}$ &  -12.9 & -12.5 & -12.2 & -12.0 & -12.0 & -12.6 & -13.4 & -13.3 & -12.6 & -11.2 \\ 
                $q_{50}$ & 0.0 & -0.0 & -0.0 & -0.0 & -0.0 & -0.0 & 0.1 & 0.5 & 0.9 & 1.3 \\ 
                $q_{84}$ & 15.5 & 15.0 & 14.7 & 14.5 & 14.9 & 15.9 & 17.3 & 17.9 & 17.9 & 17.2 \\ 
                \hline
                \multicolumn{11}{c}{C3}\\
                $q_{16}$ & -17.4 & -17.5 & -17.5 & -17.3 & -17.4 & -18.2 & -19.3 & -18.3 & -16.0 & -12.1  \\ 
                $q_{50}$ & 0.1 & 0.0 & 0.0 & 0.0 & 0.0 & -0.0 & -0.1 & -0.1 & 0.1 & 0.4  \\ 
                $q_{84}$ & 19.4 & 20.9 & 22.2 & 23.2 & 24.1 & 24.8 & 25.2 & 24.7 & 23.5 & 21.6  \\ 
                \hline
                \multicolumn{11}{c}{$\alpha_{\rm ml}$ = 1.74}\\
                $q_{16}$ & -8.7 & -8.5 & -8.2 & -8.1 & -8.0 & -8.5 & -9.3 & -9.5 & -9.1 & -8.1 \\ 
                $q_{50}$ &  4.9 & 4.8 & 4.7 & 4.6 & 4.5 & 4.6 & 4.9 & 5.2 & 5.2 & 4.9 \\ 
                $q_{84}$ & 21.3 & 20.8 & 20.5 & 20.4 & 20.9 & 22.1 & 23.7 & 24.0 & 23.1 & 21.0 \\ 
                \hline
                \multicolumn{11}{c}{no diffusion}\\
                $q_{16}$ &       & -38.8 & -37.9 & -36.8 & -35.4 & -32.8 & -28.3 & -23.1 & -17.7 & -12.2 \\ 
                $q_{50}$ &       & -29.0 & -28.2 & -26.9 & -24.8 & -20.9 & -15.2 & -9.7 & -4.3 & 0.9 \\  
                $q_{84}$ &       & -14.9 & -13.6 & -11.9 & -9.3 & -4.9 & 0.9 & 6.3 & 11.3 & 15.9 \\
                \hline
        \end{tabular}
\end{table*}

\begin{table*}[]
        \centering
        \caption{Same as Table~\ref{tab:global_MS} but for RGB stars.} 
        \label{tab:global_RGB}
        \begin{tabular}{lcccccccccc}
                \hline\hline
                \multicolumn{11}{c}{Red Giant Branch}\\
                \hline\hline
                & \multicolumn{10}{c}{Mass ($M_{\sun}$)}\\
                & 0.76 & 0.78 & 0.80 & 0.82 & 0.84 & 0.86 & 0.88 & 0.90 & 0.92 & 0.94\\
                \hline
                \multicolumn{11}{c}{C1}\\
                $q_{16}$ & -39.4 & -40.0 & -40.1 & -39.5 & -38.2 & -35.8 & -32.5 & -29.2 & -25.8 & -22.4 \\ 
                $q_{50}$ & 1.8 & 1.4 & 1.1 & 1.0 & 1.2 & 1.5 & 1.7 & 1.6 & 1.5 & 1.3 \\ 
                $q_{84}$ & 37.1 & 44.0 & 51.4 & 59.6 & 67.1 & 72.1 & 74.3 & 75.6 & 76.2 & 76.2 \\ 
                \hline
                \multicolumn{11}{c}{C2}\\
                $q_{16}$ & -19.6 & -19.3 & -19.3 & -19.3 & -19.5 & -19.9 & -20.2 & -20.1 & -19.5 & -18.5 \\ 
                $q_{50}$ &  0.0 & 0.0 & 0.0 & 0.0 & 0.0 & -0.0 & -0.0 & 0.0 & 0.0 & 0.1 \\
                $q_{84}$ & 25.5 & 25.1 & 24.8 & 24.6 & 24.7 & 25.2 & 25.9 & 26.3 & 26.6 & 26.7 \\
                \hline
                \multicolumn{11}{c}{C3}\\
                $q_{16}$ & -34.6 & -35.1 & -35.2 & -35.1 & -34.7 & -33.3 & -31.0 & -28.2 & -25.0 & -21.4 \\ 
                $q_{50}$ & 0.3 & 0.4 & 0.5 & 0.4 & 0.4 & 0.2 & 0.1 & 0.4 & 1.4 & 3.0 \\ 
                $q_{84}$ & 35.7 & 41.6 & 47.1 & 52.7 & 56.9 & 59.2 & 60.1 & 60.6 & 61.5 & 62.7 \\ 
                \hline
                \multicolumn{11}{c}{$\alpha_{\rm ml}$ = 1.74}\\
                $q_{16}$ & -21.9 & -21.9 & -21.9 & -21.8 & -22.1 & -22.6 & -23.0 & -22.8 & -22.0 & -20.6 \\ 
                $q_{50}$ &  -2.0 & -2.3 & -2.7 & -3.2 & -3.6 & -3.8 & -3.8 & -3.8 & -3.8 & -3.7 \\  
                $q_{84}$ & 20.7 & 20.4 & 20.4 & 20.7 & 21.3 & 22.1 & 22.7 & 22.8 & 22.4 & 21.6 \\
                \hline
                \multicolumn{11}{c}{no diffusion}\\
                $q_{16}$ &       & -22.5 & -22.2 & -22.0 & -21.9 & -22.2 & -22.5 & -22.6 & -22.5 & -22.2 \\
                $q_{50}$ &       & -3.6 & -3.4 & -3.2 & -3.1 & -3.0 & -2.9 & -2.9 & -3.0 & -3.3 \\ 
                $q_{84}$ &       & 20.9 & 20.8 & 20.9 & 21.2 & 21.6 & 22.1 & 22.6 & 22.5 & 21.7 \\ 
                \hline
        \end{tabular}
\end{table*}

\begin{table*}[]
        \centering
        \caption{SCEPtER median ($q_{50}$) and $1 \sigma$ envelope boundaries
                ($q_{16}$ and 
                $q_{84}$) for the relative error on
age as a
                function of the relative age of the star during the MS. We note that
                MS relative ages lower than 0.4 do not exist in our sample as they are simply too low.} 
        \label{tab:global-pcage_MS}
        \begin{tabular}{lccccccc}
            \hline\hline
                \multicolumn{8}{c}{Main Sequence}\\
                \hline\hline
                & \multicolumn{7}{c}{relative age}\\
                & 0.4 & 0.5 & 0.6 & 0.7 & 0.8 & 0.9 & 1.0\\
                \hline
                \multicolumn{8}{c}{C1}\\
                $q_{16}$   &   -29.7 & -28.1 & -25.6 & -22.6 & -19.7 & -17.2 & -15.4\\
                $q_{50}$   &   -5.9 & -4.3 & -2.8 & -1.6 & -0.7 & -0.2 & 0.0 \\ 
                $q_{84}$   &   9.6 & 11.4 & 13.3 & 15.6 & 17.4 & 18.3 & 18.5 \\ 
                \hline
                \multicolumn{8}{c}{C2}\\
                $q_{16}$   &   -25.4 & -24.4 & -22.7 & -20.5 & -18.3 & -16.2 & -14.4 \\ 
                $q_{50}$   &   -3.6 & -2.8 & -1.9 & -1.0 & -0.3 & 0.0 & -0.0 \\  
                $q_{84}$   &   4.1 & 7.6 & 11.0 & 14.1 & 16.2 & 17.0 & 16.8 \\ 
                \hline
                \multicolumn{8}{c}{C3}\\
                $q_{16}$   &  -34.5 & -34.1 & -32.9 & -30.8 & -27.9 & -24.4 & -21.0  \\ 
                $q_{50}$   &   -5.7 & -4.4 & -3.2 & -2.1 & -1.1 & -0.4 & 0.0 \\ 
                $q_{84}$   &  5.0 & 8.1 & 12.0 & 16.3 & 19.4 & 21.9 & 23.5 \\ 
                \hline
                \multicolumn{8}{c}{$\alpha_{\rm ml}$ = 1.74}\\
                $q_{16}$   &   -12.1 & -12.4 & -12.1 & -11.2 & -10.1 & -9.2 & -8.4 \\ 
                $q_{50}$   &  4.2 & 3.7 & 3.8 & 4.8 & 6.5 & 7.1 & 6.5 \\ 
                $q_{84}$   &   12.5 & 15.3 & 18.2 & 21.4 & 23.9 & 24.9 & 24.5 \\ 
                \hline
                \multicolumn{8}{c}{no diffusion}\\
                $q_{16}$   &   -66.7 & -62.5 & -57.9 & -53.6 & -50.2 & -46.5 & -42.9 \\  
                $q_{50}$   &   -45.6 & -42.9 & -39.8 & -37.1 & -35.1 & -33.1 & -31.2 \\ 
                $q_{84}$   &   -23.1 & -20.7 & -19.0 & -18.4 & -18.0 & -16.9 & -15.5 \\  
                \hline
        \end{tabular}
\end{table*}

\begin{table*}[]
        \centering
        \label{tab:global-pcage_SGB}
        \caption{Same as Table~\ref{tab:global-pcage_MS} but for SGB stars.}
        \begin{tabular}{lcccccccccc}
            \hline\hline
                \multicolumn{11}{c}{Sub-Giant Branch}\\
                \hline\hline
                & \multicolumn{10}{c}{relative age}\\
                & 0.1 & 0.2 & 0.3 & 0.4 & 0.5 & 0.6 & 0.7 & 0.8 & 0.9 & 1.0\\
                \hline
                \multicolumn{11}{c}{C1}\\
                $q_{16}$ & -13.2 & -12.9 & -12.6 & -12.4 & -12.3 & -12.2 & -12.4 & -13.3 & -14.7 & -16.8 \\ 
                $q_{50}$ & -0 & 0.1 & 0.1 & 0.1 & 0.1 & 0 & -0.1 & -0.2 & -0.3 & -0.5\\ 
                $q_{84}$ & 15.9 & 15.6 & 15.3 & 15.0 & 14.9 & 14.8 & 15.1 & 16.7 & 19.9 & 24.8 \\ 
                \hline
                \multicolumn{11}{c}{C2}\\
                $q_{16}$  & -12.7 & -12.4 & -12.2 & -12.0 & -11.8 & -11.7 & -11.8 & -12.4 & -13.3 & -14.7 \\ 
                $q_{50}$ & 0 & 0 & 0 & 0 & 0 & 0 & -0 & -0 & 0 & 0 \\
                $q_{84}$ & 15.5 & 15.2 & 15.0 & 14.8 & 14.7 & 14.4 & 14.5 & 15.2 & 16.7 & 19.1 \\ 
                \hline
                \multicolumn{11}{c}{C3}\\
                $q_{16}$ &-18.2 & -17.5 & -17.0 & -16.6 & -16.3 & -16.0 & -16.3 & -17.3 & -19.0 & -21.5  \\ 
                $q_{50}$ & 0 & 0.1 & 0.1 & 0 & 0 & 0 & 0 & -0 & -0.1 & -0.3\\ 
                $q_{84}$ & 22.2 & 21.9 & 21.6 & 21.4 & 21.1 & 20.4 & 20.3 & 22.1 & 26.1 & 32.5 \\ 
                \hline
                \multicolumn{11}{c}{$\alpha_{\rm ml}$ = 1.74}\\
                $q_{16}$ &  -8.0 & -7.9 & -7.9 & -7.9 & -8.0 & -7.9 & -8.1 & -8.8 & -9.9 & -11.5 \\  
                $q_{50}$ &   5.3 & 5.2 & 5.1 & 4.9 & 4.8 & 4.7 & 4.3 & 3.9 & 3.6 & 3.4 \\ 
                $q_{84}$ &  22.1 & 21.7 & 21.3 & 20.8 & 20.4 & 20.0 & 19.8 & 20.2 & 21.3 & 23.2 \\
                \hline
                \multicolumn{11}{c}{no diffusion}\\
                $q_{16}$ &  -40.4 & -39.8 & -39.1 & -38.3 & -37.4 & -36.2 & -34.2 & -31.4 & -27.8 & -23.3 \\  
                $q_{50}$ & -29.9 & -29.6 & -29.1 & -28.5 & -27.7 & -26.6 & -24.4 & -20.8 & -16.1 & -10.1 \\
                $q_{84}$ & -16.2 & -15.9 & -15.6 & -15.1 & -14.3 & -13.0 & -10.4 & -6.3 & -0.6 & 6.7 \\  
                \hline
        \end{tabular}
\end{table*}

\begin{table*}[]
        \centering
        \caption{Same as Table~\ref{tab:global-pcage_MS} but for RGB stars.} 
        \label{tab:global-pcage_RGB}
        \begin{tabular}{lcccccccccc}
            \hline\hline
                \multicolumn{11}{c}{Red Giant Branch}\\
                \hline\hline
                & \multicolumn{10}{c}{relative age}\\
                & 0.1 & 0.2 & 0.3 & 0.4 & 0.5 & 0.6 & 0.7 & 0.8 & 0.9 & 1.0\\
                \hline
                \multicolumn{11}{c}{C1}\\
                $q_{16}$ & -29.4 & -33.8 & -37.2 & -39.7 & -40.6 & -40.6 & -40.2 & -39.5 & -38.6 & -37.4 \\ 
                $q_{50}$ & 2.2 & 2.1 & 1.6 & 1.0 & 0.6 & 0.2 & 0 & 0.2 & 1.0 & 2.2\\ 
                $q_{84}$ & 48.0 & 51.2 & 53.6 & 55.2 & 56.1 & 56.7 & 57.1 & 57.3 & 57.0 & 56.2 \\ 
                \hline
                \multicolumn{11}{c}{C2}\\
                $q_{16}$ & -19.2 & -19.7 & -20.1 & -20.3 & -20.2 & -20.1 & -19.8 & -19.4 & -18.7 & -17.8 \\ 
                $q_{50}$ & 0.1 & 0.1 & 0.1 & 0.0 & -0.0 & -0.0 & -0.0 & -0.0 & -0.0 & 0.0 \\ 
                $q_{84}$ &  25.9 & 25.8 & 25.5 & 25.0 & 24.6 & 24.6 & 24.6 & 24.5 & 24.3 & 24.1 \\ 
                \hline
                \multicolumn{11}{c}{C3}\\
                $q_{16}$ & -31.0 & -33.3 & -34.8 & -35.7 & -35.8 & -35.4 & -34.8 & -33.8 & -32.5 & -30.7 \\ 
                $q_{50}$ & 2.5 & 2.1 & 1.5 & 0.6 & 0 & -0 & 0 & 0 & 0 & 0 \\ 
                $q_{84}$ &  47.7 & 48.7 & 49.1 & 48.8 & 48.1 & 47.2 & 46.8 & 46.6 & 46.4 & 46.2\\ 
                \hline
                \multicolumn{11}{c}{$\alpha_{\rm ml}$ = 1.74}\\
                $q_{16}$ & -17.9 & -20.0 & -21.6 & -22.6 & -23.0 & -23.1 & -23.3 & -23.5 & -23.8 & -24.0 \\ 
                $q_{50}$ &1.0 & -0.6 & -2.0 & -3.2 & -4.0 & -4.4 & -4.9 & -5.1 & -5.0 & -4.3 \\ 
                $q_{84}$ & 25.1 & 23.7 & 22.2 & 20.7 & 19.7 & 19.2 & 18.8 & 18.8 & 19.4 & 20.6 \\
                \hline
                \multicolumn{11}{c}{no diffusion}\\
                $q_{16}$  & -24.2 & -23.2 & -22.3 & -21.9 & -21.4 & -21.3 & -21.6 & -21.6 & -21.5 & -21.2 \\ 
                $q_{50}$  & -4.5 & -3.4 & -2.9 & -2.7 & -2.8 & -3.0 & -3.1 & -3.1 & -3.0 & -2.9 \\ 
                $q_{84}$  & 20.5 & 21.8 & 22.2 & 21.9 & 21.2 & 20.8 & 20.8 & 21.1 & 21.6 & 22.2 \\ 
                \hline
        \end{tabular}
\end{table*}

\begin{figure*}
\centering
\includegraphics[width=6cm]{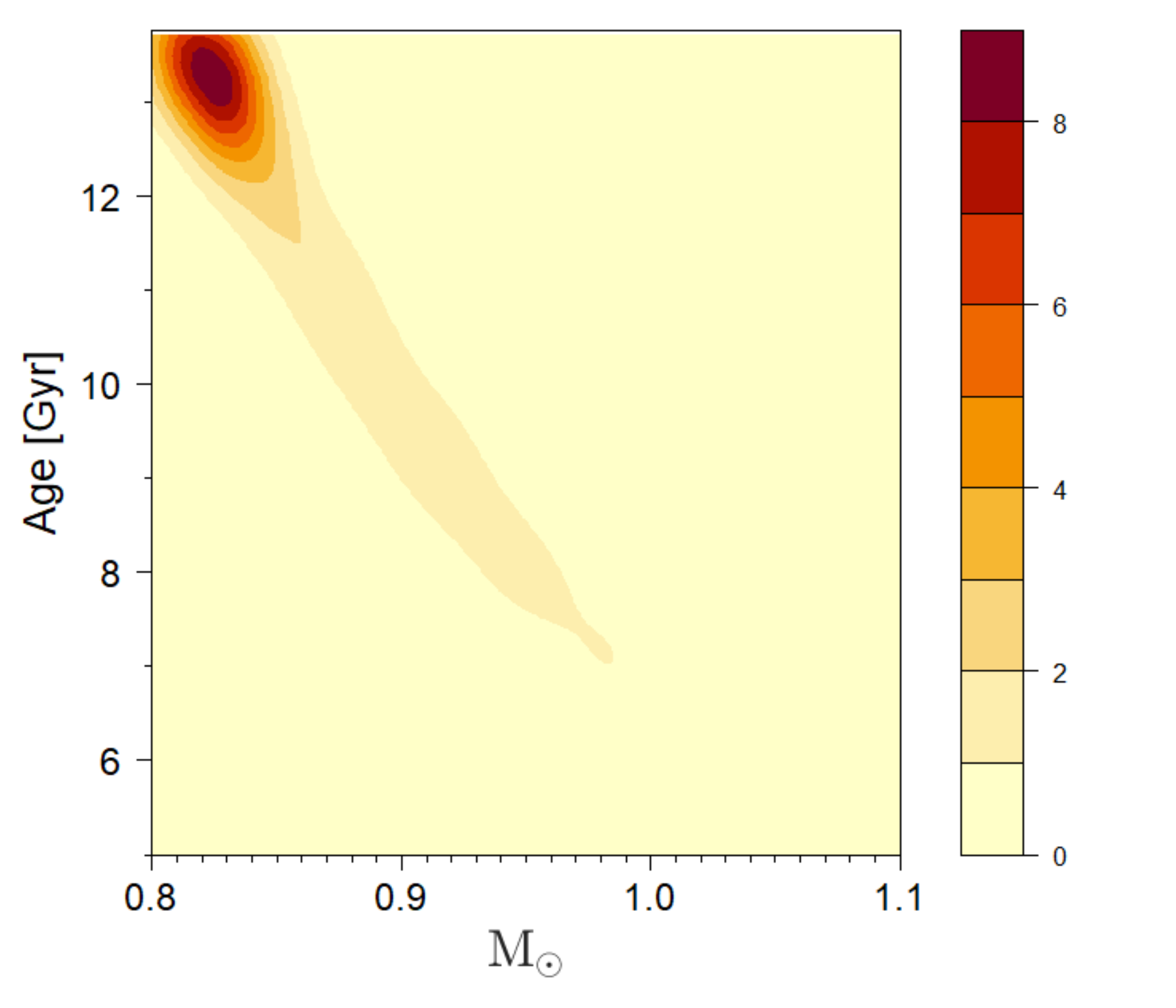}
\includegraphics[width=6cm]{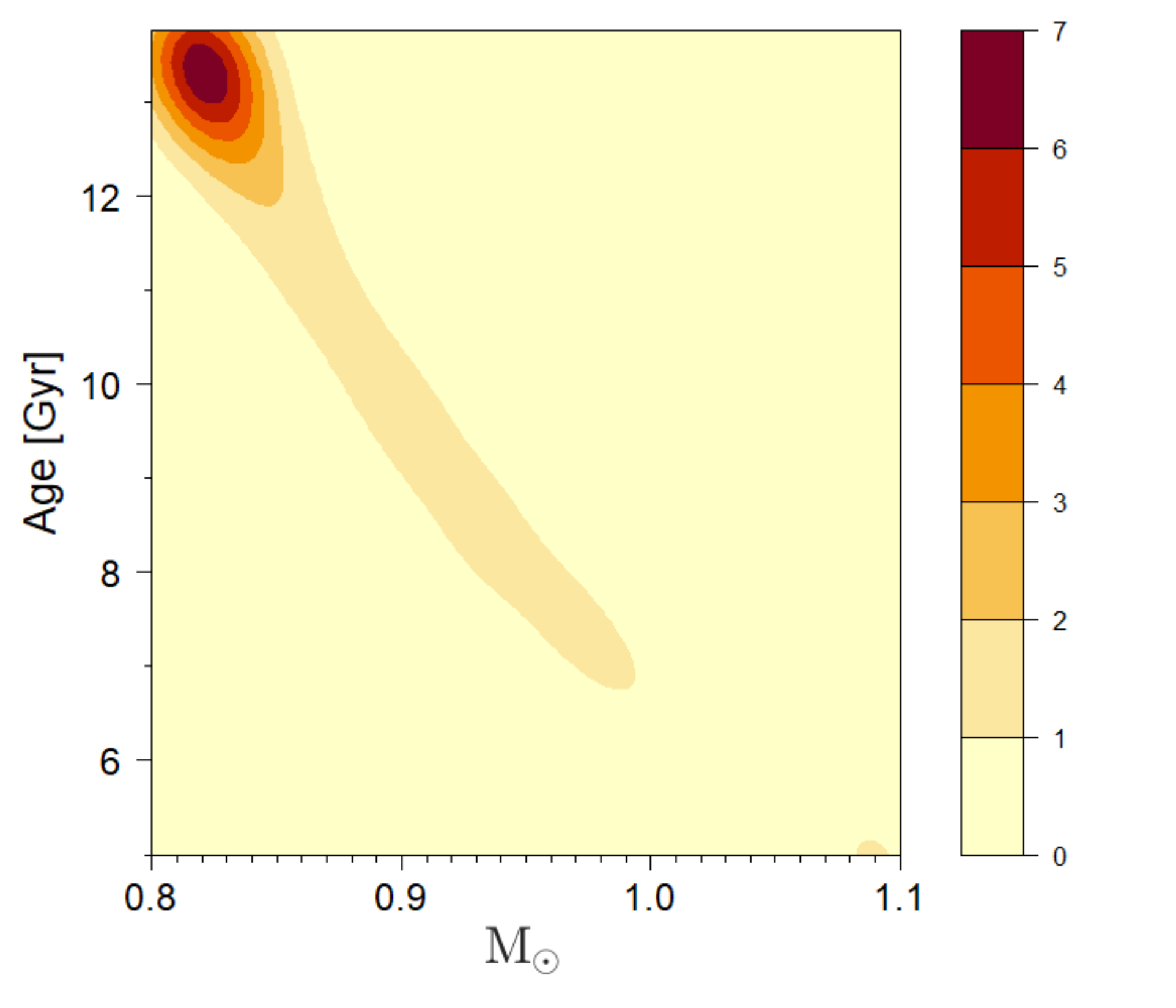}
\includegraphics[width=6cm]{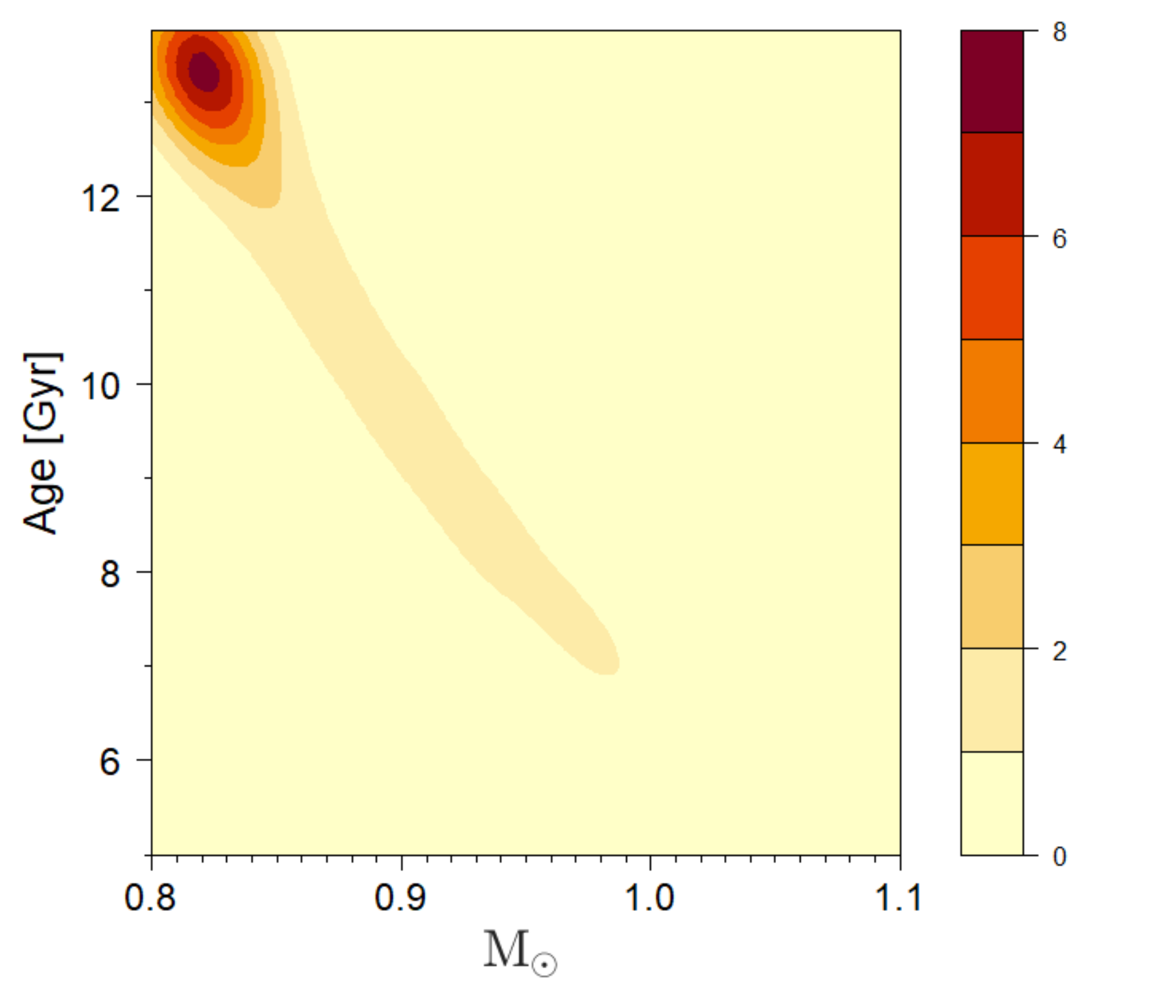}

\includegraphics[width=6cm]{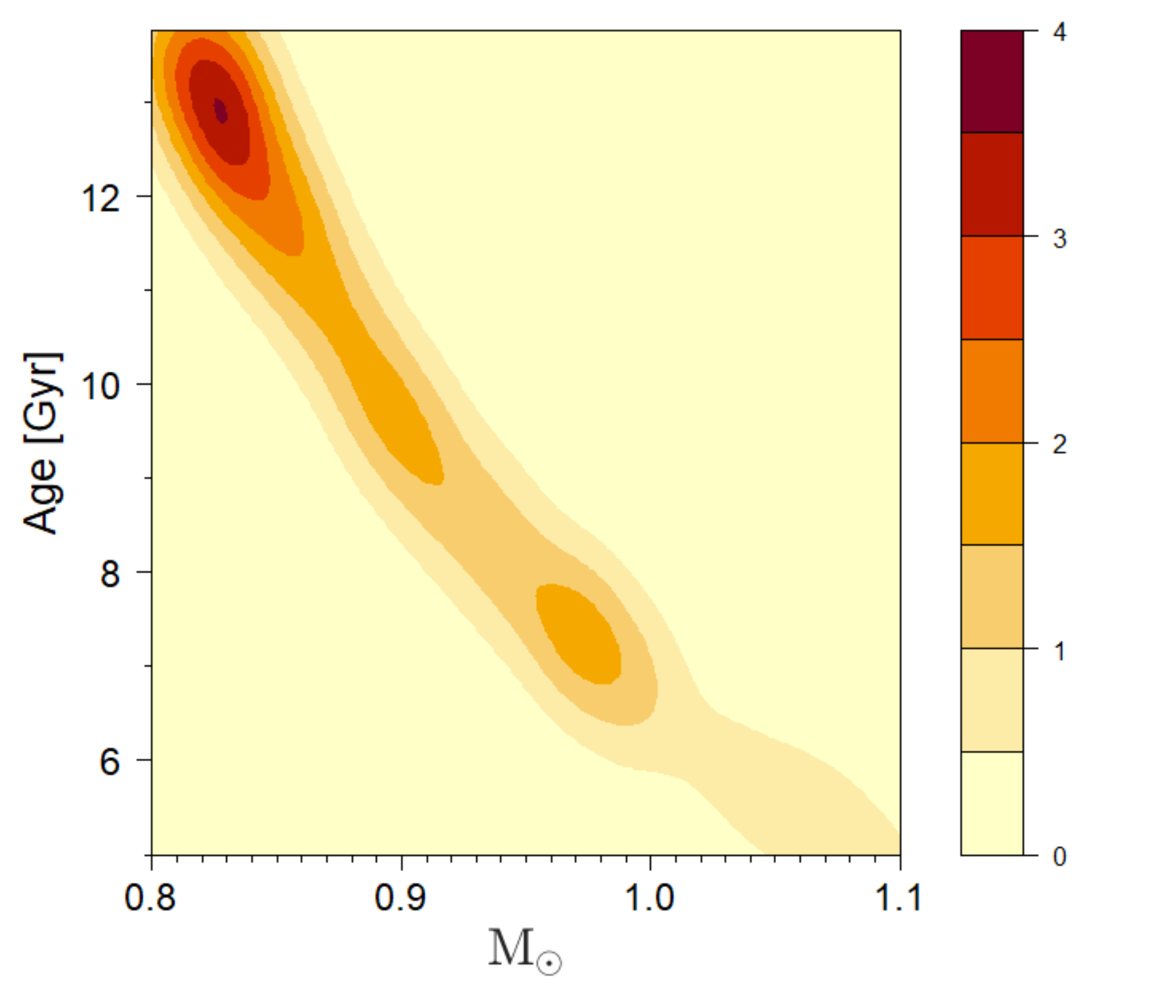}
\includegraphics[width=6cm]{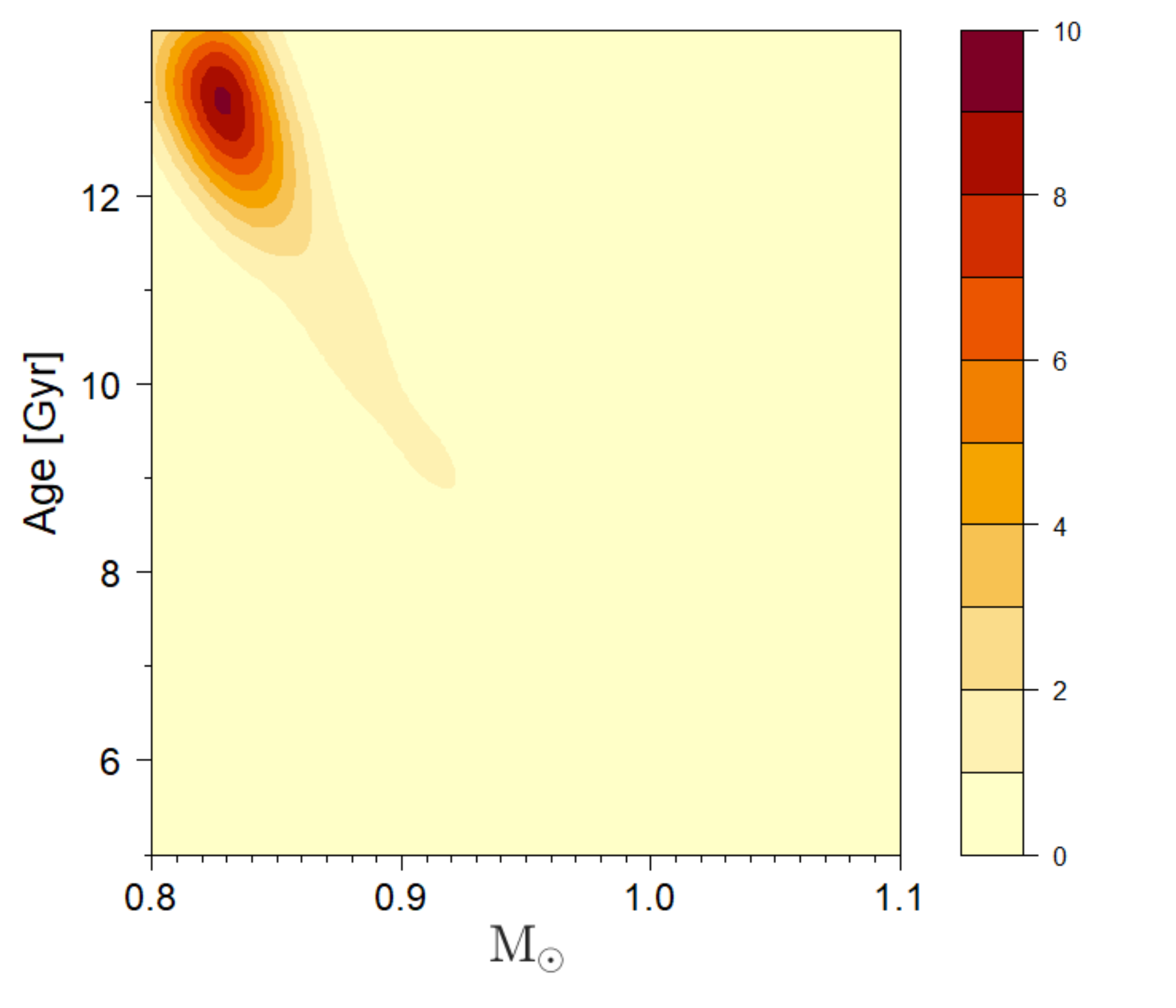}
\includegraphics[width=6cm]{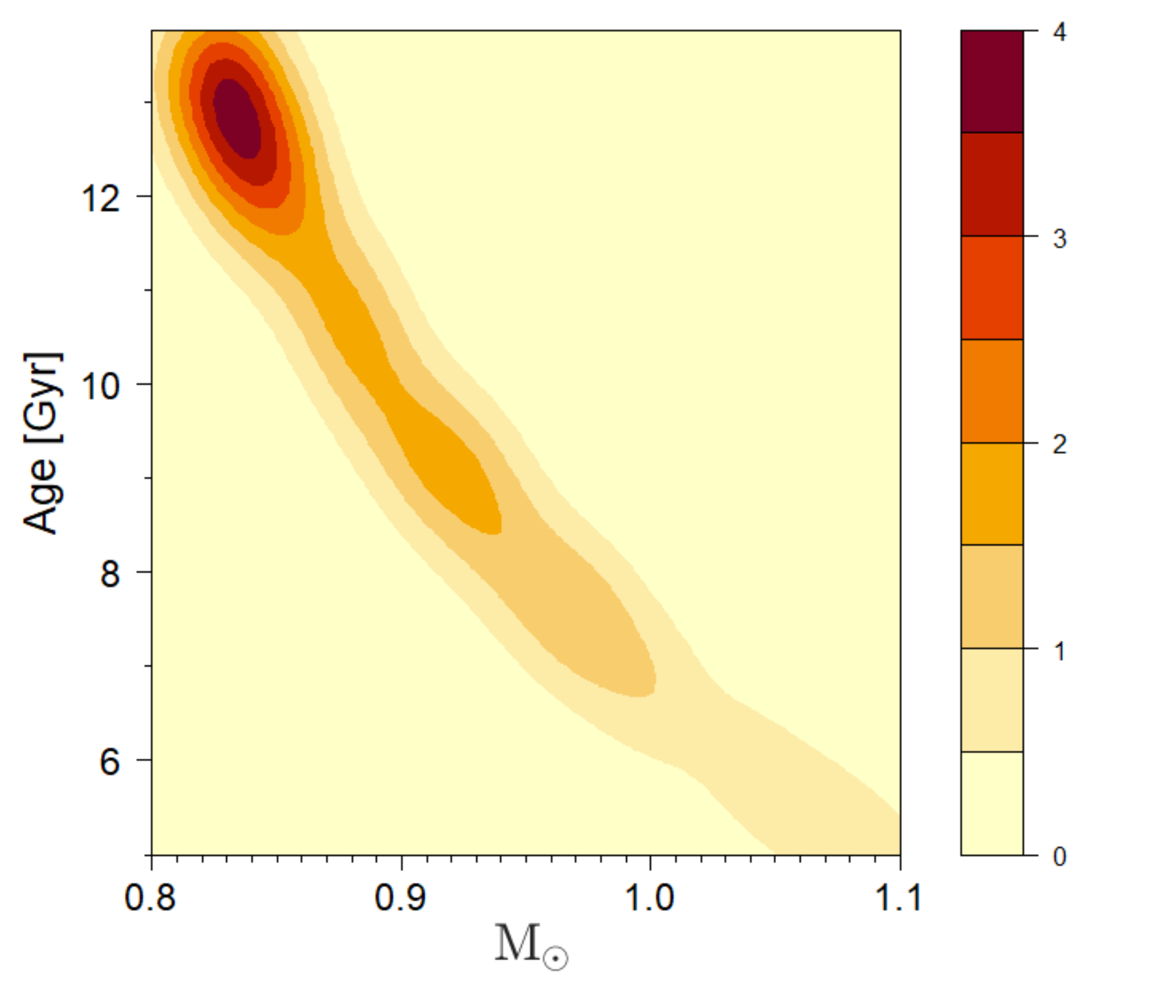}
\caption{{Top row}: Maximum-likelihood 2D density plots in the age vs mass plane for stars S1, S2, and S3 in M4. {\it Bottom row}: Same as in the top row for stars S5, S6, and S7.}
\label{fig:M4results}
\end{figure*}


\end{document}